\documentclass[12pt]{article}
\usepackage{latexsym,amsfonts,amsmath,amssymb,theorem,dsfont,wasysym}
\usepackage{fullpage}

\DeclareFontFamily{OT1}{rsfs10}{}
\DeclareFontShape{OT1}{rsfs10}{m}{n}{ <-> rsfs10 }{}
\DeclareMathAlphabet{\mathscript}{OT1}{rsfs10}{m}{n}

\def\gt{\rightarrow}
\def\be#1\ee{\begin{align}#1\end{align}}
\def\del{\partial}
\def\grad{\nabla}
\def\({\left(}
\def\){\right)}

\def\nn{\nonumber}
\def\ns{\normalsize}

\def\td{\tilde{\delta}}

\def\dt{\mathrm{d} t \,}

\def\dcx{\mathrm{d}^3 x \,}
\def\dcy{\mathrm{d}^3 y \,}
\def\dcz{\mathrm{d}^3 z \,}

\def\H{\mathcal{H}}
\def\I{\mathcal{I}}
\def\J{\mathcal{J}}
\def\K{\mathcal{K}}
\def\L{\Lambda}
\def\M{\mathcal{M}}
\def\N{\mathcal{N}}

\def\gradt{\tilde{\nabla}}
\def\Gt{\tilde{\Gamma}}
\def\Ht{\tilde{\H}}
\def\Nt{\tilde{N}}
\def\Rt{\tilde{R}}
\def\Rb{\bar{R}}

\def\pt{\tilde{\pi}}
\def\th{\tilde{h}}
\def\tg{\tilde{g}}
\def\pw{\pi_{\omega}}
\def\pgr{\pi_{\rm GR}}
\def\Pw{\Pi_{\omega}}

\def\pb{\bar{\phi}}

\def\a{\alpha}

\def\d{\delta}

\def\k{\kappa}
\def\l{\lambda}

\def\s{\sigma}

\def\f{\varphi}

\def\w{\omega}

\def\G{\Gamma}
\def\D{\Delta}

\def\W{\Omega}

\begin{document}

\begin{titlepage}

\title{
  \hfill{\ns }  \\
{\LARGE Spatially Covariant Theories of a Transverse, Traceless Graviton, Part I: Formalism}
\\
}
\author{
Justin Khoury$^1$, Godfrey E. J. Miller$^1$ and Andrew J. Tolley$^2$
\\
{\ns ${}^1$ Center for Particle Cosmology, Department of Physics \& Astronomy} \\[-0.25cm]
{\ns University of Pennsylvania, 209 South 33rd Street, Philadelphia, PA 19104}\\
{\ns ${}^2$ Department of Physics, Case Western Reserve University} \\[-0.25cm]
{\ns 10900 Euclid Ave, Cleveland, OH 44106}\\
[0.3cm]}

\date{}

\maketitle

\begin{abstract}
General relativity is a covariant theory of two transverse, traceless graviton degrees of freedom.  According to a theorem of Hojman, Kucha\v{r}, and Teitelboim, modifications of general relativity must either introduce new degrees of freedom or violate the principle of general covariance.  In this paper, we explore modifications of general relativity that retain the same number of gravitational degrees of freedom, and therefore explicitly break general covariance.  Motivated by cosmology, the modifications of interest maintain spatial covariance. Demanding consistency of the theory forces the physical Hamiltonian density to obey an analogue of the renormalization group equation.  In this context, the equation encodes the invariance of the theory under flow through the space of conformally equivalent spatial metrics.  This paper is dedicated to setting up the formalism of our approach and applying it to a realistic class of theories.  Forthcoming work will apply the formalism more generally.
\end{abstract}

\end{titlepage}

\section{Introduction}
For nearly a century, general relativity has been the most successful paradigm for interpreting and understanding classical gravitational phenomena, and to this day there have been no unequivocal refutations of general relativity.  Nonetheless, there are compelling reasons to study alternative gravitational theories.

Perhaps the most obvious reason is to explain {\it empirical} anomalies, most notably the observed cosmic acceleration.  While it is true that this phenomenon can be understood in terms of a cosmological constant in pure general relativity, this approach has the drawback that we have no handle on what physics might set the magnitude of the cosmological constant.  It is currently an outstanding theoretical challenge to determine what physical degrees of freedom are associated with late-time acceleration.  Dynamical theories of dark energy postulate scalar degrees of freedom similar to those invoked to account for inflation.  Unfortunately, there is no unambiguous evidence for such scalar degrees of freedom.  One motivation for this paper is the possibility that cosmic acceleration might be directly associated with the graviton degrees of freedom.

Apart from any attempt to understand empirical anomalies, there remains a compelling {\it theoretical} reason to study alternatives: to determine which features of general relativity are essential to its experimental success, and which features are merely incidental.  To analyze the theory in this manner, we must know what freedom we have to modify the theory while retaining its explanatory power.

The two transverse, traceless graviton degrees of freedom are a key feature of general relativity.  Though graviton exchange has never been measured and gravitational waves have never been directly detected, there is substantial {\it indirect} evidence for the existence of these degrees of freedom.  It is logically possible that additional gravitational degrees of freedom exist, but at the same time there is no unambiguous evidence for them.  It is therefore natural to ask whether and how we can modify general relativity while preserving the same number of degrees of freedom.

In this paper, we construct manifestly consistent modified theories of gravity that retain the same local degrees of freedom as general relativity.  To evade the consequences of the theorem that general relativity is the unique theory of a massless spin-2 particle~\cite{Feynman:1996kb,Hinterbichler:2011tt}, our theories break local Lorentz symmetry {\it explicitly}.  Theories in which Lorentz symmetry is only broken spontaneously necessarily rely on additional local degrees of freedom. These appear in the broken phase as massless Goldstone modes; an example of such a theory is ghost condensation~\cite{ArkaniHamed:2003uy}.

Perhaps the biggest obstacle to modifying general relativity is the opaqueness of the theory.  General relativity as formulated by Einstein and Hilbert is a {\it covariant} theory, which means that the equations of motion for the spacetime metric $g_{\mu \nu}$ take the same form in any coordinate system.  Unfortunately, invariance under coordinate transformations implies that the theory contains a great deal of gauge arbitrariness, and the underlying dynamical degrees of freedom of the theory have proven difficult to isolate.  In fact, the notorious elusiveness of the physical degrees of freedom is partially responsible for the difficulty of quantizing general relativity.

This gauge arbitrariness can be understood most clearly by treating general relativity as a {\it constrained field theory}.  By writing the metric $g_{\mu \nu}$ in ADM form\footnote{{\it i.e.}, in terms of a spatial metric $h_{i j}$, a lapse $N \equiv N^0$, and a shift $N^i$.} and discarding a boundary term, the Einstein-Hilbert action can be rewritten in canonical form as a theory of a spatial metric $h_{i j}$ and a conjugate momentum tensor $\pi^{i j}$ subject to four {\it first class} constraints $\H_{\mu}$~\cite{Arnowitt:1962hi}.  While this form of the theory is not written in terms of manifestly diffeomorphism covariant objects, the covariance of the theory can still be inferred from the covariance ``algebra'' satisfied by the $\H_{\mu}$, which act as the generators of spacetime diffeomorphisms under the action of the Poisson bracket~\cite{Teitelboim:1972vw}.  By representing the gauge symmetries of general relativity as constraints on the phase space of the theory, it becomes straightforward to count degrees of freedom.  According to the standard counting prescription, it follows from the presence of four first class constraints $\H_{\mu}$ in a theory of six canonical coordinates $h_{i j}$ that general relativity contains two local degrees of freedom.\footnote{See section~\ref{gr4dof} for more detail.}  In the passage to quantum theory, these transverse, traceless degrees of freedom become the two polarizations of the graviton.

To isolate the physical graviton degrees of freedom, one would have to solve the four constraints $\H_{\mu}$.  By taking the configuration space for the metric to be Wheeler's superspace, it is possible to solve the three momentum constraints $\H_i$ by fiat, but the Hamiltonian constraint $\H_0$ has thus far defied solution in general.  Unless the Hamiltonian constraint can be solved, the gauge-arbitrariness of general relativity cannot be eliminated.  Fortunately, though no general solution to the Hamiltonian constraint has been found, it is possible to solve it in certain circumstances by imposing an appropriate gauge-fixing condition.

In 1974, Hojman, Kucha\v{r}, and Teitelboim (HKT) proved that general relativity with a cosmological constant is the unique covariant theory of a spatial metric $h_{i j}$ and its conjugate momentum $\pi^{i j}$; in fact, general relativity is the minimal representation of the covariance algebra~\cite{Hojman:1976vp,Kuchar:1974es}.  It follows that alternative {\it covariant} theories of gravitation must introduce additional degrees of freedom beyond the two graviton degrees of freedom in general relativity~\cite{Anderson:2005am}.  Conversely, alternative theories of a spatial metric $h_{i j}$ and its conjugate momentum $\pi^{i j}$ cannot be covariant.  To modify general relativity, one must either introduce new degrees of freedom or violate the principle of general covariance.

We wish our theories to retain the same local degrees of freedom as general relativity, so in accordance with the theorem of HKT, our theories cannot be diffeomorphism covariant.  This aspect of our approach is not necessarily a defect.  The covariance property of general relativity may be intellectually appealing, but the form invariance of the equations of motion is purchased at the cost of substantial gauge redundancy.  Moreover, since we do not observe exact spacetime symmetry in our universe, this property of general relativity is not necessarily key to the success of the theory.  Simply put, on cosmological scales there is a strong asymmetry between the past and the future, and the observable universe has a preferred rest frame; these observations are conventionally understood as a result of spontaneous symmetry breaking, but {\it explicit} symmetry breaking is another logical possibility.

That being said, on cosmological scales in the cosmological rest frame there is substantial evidence for spatial homogeneity and isotropy.  To maximize the verisimilitude of our treatment, the theories we consider will retain explicit covariance under {\it spatial} diffeomorphisms.  To summarize, we will attempt to modify general relativity while preserving 1) the number of graviton degrees of freedom, and 2) covariance under spatial diffeomorphisms.  In this paper, we develop a general framework within which to explore the freedom we have to modify general relativity while retaining these two desirable properties.

Concretely, we will begin by recasting general relativity in spatially covariant form, by solving the Hamiltonian constraint while preserving the momentum constraints.  We will solve the Hamiltonian constraint by imposing a cosmologically motivated gauge constraint: we will take the determinant of the spatial metric to be the measure of time.  This operation destroys the manifest diffeomorphism covariance and local Lorentz covariance of the theory.  We emphasize that this gauge breaks down in the general case when the determinant of the spatial metric is allowed to evolve non-monotonically, but it is a natural choice when considering perturbative corrections to FRW spacetime.  By solving the Hamiltonian constraint, the determinant of the spatial metric and the trace of the momentum tensor drop out of the phase space of the theory.  We thereby obtain general relativity as a theory of a unit-determinant metric $\th_{i j}$ and a traceless conjugate momentum tensor $\pt^{i j}$ subject to three first class momentum constraints $\Ht_i$, which act as the generators of spatial diffeomorphisms.  By the standard counting prescription, the presence of three first class constraints $\Ht_i$ in a theory of five canonical coordinates $\th_{i j}$ guarantees that spatially covariant general relativity contains two degrees of freedom, as it should.\footnote{See section~\ref{gr3dof} for more detail.}

Our strategy for modifying general relativity relies on the fact that {\it any} theory of five canonical coordinates subject to three first class constraints contains two degrees of freedom.  To modify general relativity, we will modify the functional form of the physical Hamiltonian density on the reduced phase space ($\th_{i j}$, $\pt^{i j}$), subject to the condition that the momentum constraints $\Ht_i$ remain first class; to ensure the consistency of the modification, we will also demand that the constraints $\Ht_i$ remain preserved by the equations of motion.  Any theory that satisfies these two restrictions will retain manifest spatial covariance, and by the counting prescription will necessarily contain two graviton degrees of freedom.  In this paper, we introduce the formalism necessary to pursue this program of modification and apply the formalism to a class of realistic theories.  Forthcoming work will apply the formalism developed here to a broader class of theories~\cite{godfreytoappear}.

The literature abounds with many and varied approaches to the pursuit of modified gravity theories, but covariant modifications of general relativity that introduce additional degrees of freedom have been the most widely explored.  The well-known method for finding covariant theories is to construct a diffeomorphism-invariant Lagrangian density out of manifestly covariant objects by contracting all free spacetime indices.  Using this technique, all manner of theories have been explored: scalar-tensor theories~\cite{Brans:1961sx}, theories with higher-order curvature terms~\cite{fR,f(R)2,f(R)3}, 
theories of massive gravity~\cite{vainshtein,ags,deRham:2010gu,deRham:2010ik,deRham:2010kj,Hassan:2011hr}, higher-dimensional gravity theories~\cite{Dvali:2000hr,dvali2,deRham:2007xp,deRham:2007rw}, galileons~\cite{ddgv,lpr,Nicolis:2008in}, chameleons~\cite{Khoury:2003aq,Khoury:2003rn,Gubser:2004uf,Brax:2004qh}, symmetrons~\cite{Hinterbichler:2010es,Hinterbichler:2011ca}, {\it etc.} For a comprehensive review of Lorentz-invariant massive gravity theories with detailed references, see~\cite{Hinterbichler:2011tt}. For a comprehensive review of observational tests of modified gravity, see~\cite{Jain:2010ka}.

Non-covariant approaches have been tried as well, but there is no single unifying procedure for the construction of such theories.  In general, the natural procedure for understanding non-covariant theories depends on which symmetries survive in the theory.  For example, in~\cite{Dubovsky:2004sg} Lorentz-violating massive graviton theories were classified by assuming the graviton mass to be invariant under the three-dimensional Euclidean group.  A prominent recent example of a non-covariant metric theory is Ho\v{r}ava-Lifshitz gravity~\cite{Horava:2009uw,Henneaux:2009zb,Horava:2010zj},\footnote{The original incarnation~\cite{Horava:2009uw} of Ho\v{r}ava-Lifshitz gravity struggled with consistency issues~\cite{Henneaux:2009zb} which were resolved in~\cite{Horava:2010zj} by imposing a consistent constraint algebra.} in which the phase space constraints are chosen to satisfy a non-relativistic version of the covariance algebra.  Also of note is the work of Barbour and collaborators on theories of conformally equivalent spatial metrics~\cite{Barbour:2011dn}.

This paper is organized as follows.  In section~\ref{gr4}, we cover the basic concepts of constrained field theory in the context of analyzing the phase space and constraint structure of general relativity.  In section~\ref{gr3}, we show how to impose our cosmological gauge condition and solve the Hamiltonian constraint to obtain a consistent spatially covariant formulation of general relativity.  In section~\ref{ultralocal}, we introduce the formalism of our approach to modifying gravity in the context of theories with an ultralocal physical Hamiltonian density.  In section~\ref{realistic}, we apply our method to derive consistency relations for a class of realistic theories which includes general relativity.

\section{General Relativity as a Constrained Field Theory}
\label{gr4}
In this section, we will analyze general relativity by treating it as a constrained field theory.  In particular, we will examine its phase space and constraint structure, and count its local degrees of freedom.

Our starting point is the Einstein-Hilbert action with a cosmological constant,
\be
S = \int \dt \dcx \sqrt{-g} \( R^{(4)} - 2 \L \)  .
\ee
From this action, the general covariance of the theory is manifest, but the counting of degrees of freedom is not.  The metric tensor $g_{\mu \nu}$ has ten components, but the theory has only two independent local degrees of freedom.  To facilitate the counting of degrees of freedom, it is conceptually simplest to rewrite the action in a manner which makes the counting manifest, {\it i.e.}, canonical form.  To this end, the metric $g_{\mu \nu}$ must first be expressed in ADM form, in terms of a lapse $N \equiv N^0$, a shift $N^{i}$, and a spatial metric $h_{i j}$:
\be
{\rm d}s^2 = g_{\mu \nu} {\rm d}x^{\mu} {\rm d}x^{\nu} = - N^2 {\rm d}t^2 + h_{i j} ({\rm d}x^i + N^i {\rm d}t) ({\rm d}x^j + N^j {\rm d}t) \, .
\ee
Up to a boundary term, the action of general relativity is
\be
S = \int \dt \dcx \sqrt{h} N \( K_{i j} K^{i j} - K^2 + R - 2 \L \) .
\ee
In this expression, indices are lowered with $h_{i j}$ and raised with its inverse $h^{i j}$, $R \equiv R^{(3)}$ is the Ricci scalar of the metric $h_{i j}$, the extrinsic curvature tensor $K_{i j}$ is defined by
\be
K_{i j} \equiv \frac{1}{2} N^{-1} \( \dot{h}_{i j} - \grad_i N_j - \grad_j N_i \) ,
\ee
$K \equiv h^{i j} K_{i j}$, and $\grad_i \equiv \grad^{(3)}_i$ is the covariant spatial derivative with respect to the metric $h_{i j}$.  To obtain the canonical action, one must first define the momentum conjugate to the spatial metric
\be
\pi^{ij} \equiv \frac{\d L}{\d \dot{h}_{ij}} = \sqrt{h} \( K^{i j} - K h^{i j} \) ;
\ee
the momentum $\pi^{i j}$ is three-tensor density of unit weight.\footnote{According to the standard convention, the weight of a tensor density is the number of times $\sqrt{h}$ multiplies the underlying tensor.}  By inverting the relation between $\pi^{i j}$ and $K^{i j}$, one can rewrite the action of general relativity in canonical form as
\be
S = \int \dt \dcx \( \pi^{i j}\dot{h}_{i j} - N^{\mu} \H_{\mu} \) ,
\ee
where 
\be
\H_0 &\equiv -\sqrt{h} (R - 2 \L)  + \frac{1}{\sqrt{h}} \( \pi^{ij}\pi_{ij} - \frac{1}{2}(\pi^i_{\ i})^2 \) \, , \nn \\
\H_i &\equiv -2 h_{ij} \grad_{k} \pi^{jk} \, .
\ee
Variation of the action with respect to $h_{i j}$ and $\pi^{i j}$ yields Hamilton's equations,
\be
\dot{h}_{i j}(x) = \frac{\d H}{\d \pi^{i j}(x)} \, , \qquad \dot{\pi}^{i j}(x) = -\frac{\d H}{\d h_{i j}(x)} \, ,
\ee
where the Hamiltonian $H$ is
\be
H = \int \dcx N^{\mu} \H_{\mu} \, . \label{4ham}
\ee
To evaluate the above variational derivatives, one must use the relations
\be
\frac{\d h_{i j}(x)}{\d h_{k l}(y)} = \frac{\d \pi^{k l}(x)}{\d \pi^{i j}(y)} = \d_{i j}^{k l} \d^3(x-y) \, ,
\ee
where
\be
\d_{i j}^{k l} \equiv \frac{1}{2} \( \d_i^k \d_j^l + \d_i^l \d_j^k \) .
\ee
Defining the Poisson bracket
\be
\{ A, B \} \equiv \int \dcz \( \frac{\d A}{\d h_{m n}(z)} \frac{\d B}{\d \pi^{m n}(z)} - \frac{\d A}{\d \pi^{m n}(z)} \frac{\d B}{\d h_{m n}(z)}\), \label{pbgr}
\ee
the equation of motion for any quantity $A(h_{i j}, \pi^{i j}, t)$ can be written as
\be
\dot{A}
&= \frac{\del A}{\del t} + \{ A, H \} \nn \\
&= \frac{\del A}{\del t} + \int \dcy N^{\nu}(y)  \,  \{ A, \H_{\nu}(y) \} \, .
\ee
If $A$ has no explicit dependence on time, its evolution is {\it generated} by its Poisson bracket with the $\H_{\mu}$.

Variation of the action with respect to $N^{\mu}$ yields the four constraints
\be
\H_{\mu} \sim 0 \, . \label{4con}
\ee
The symbol $\sim$ denotes {\it weak equality}, or equality after the constraints $\H_{\mu} \sim 0$ have been enforced.  For example, if $X = Y + \l^{\mu} \H_{\mu}$, then $X \sim Y$.  Since the constraints define a surface in phase space, weak equality is also termed equality {\it on the constraint surface}.  As an aside, it follows from~(\ref{4ham}) and~(\ref{4con}) that $H \sim 0$; the vanishing of the Hamiltonian on the constraint surface is a feature common to covariant theories.

There is no $\pi_{\mu} \dot{N}^{\mu}$ term that would allow us to compute a variational expression for $\dot{N}^{\mu}$, so the time evolution of $N^{\mu}$ is unconstrained by the action.  The four functions $N^{\mu}$ are thus arbitrary until and unless we gauge-fix them.

\subsection{Constraint Properties \& Degrees of Freedom}
\label{gr4dof}
Before examining the constraints more closely, we pause to review some terminology first introduced by Dirac for describing constrained theories.  A quantity whose Poisson bracket with each of the constraints vanishes (identically or weakly) is termed {\it first class}; a quantity whose Poisson bracket fails to vanish weakly with at least one constraint is termed {\it second class}.  A {\it first class constraint} has vanishing Poisson bracket with all constraints, while a {\it second class constraint} has non-vanishing Poisson bracket with at least one other constraint.  In most cases of interest, first class constraints generate {\it gauge symmetries} under the action of the Poisson bracket.  Second class constraints can usually be solved, either implicitly (by using the ``Dirac bracket'') or explicitly (by expressing some phase space variables in terms of others).

In general relativity, the constraints $\H_{\mu}$ generate spacetime diffeomorphisms.  By direct calculation --- see appendix A for details --- it is possible to prove that the constraints $\H_{\mu}$ are first class, $\{\H_{\mu}(x), \H_{\nu}(y)\} \sim 0$.  This means that the symmetry generators are closed under the action of the Poisson bracket, as they should be in order to consistently represent any kind of a symmetry.  In particular,
\be
\{\H_0 (x), \H_0 (y)\}
&= \H^i (x) \del_{x^i} \d^3(x-y) - \H^i (y) \del_{y^i} \d^3(x-y) \, , \nn \\
\{\H_0 (x), \H_i (y)\}
&= \H_0 (y) \del_{x^i} \d^3(x-y) \, , \nn \\
\{\H_i (x), \H_j (y)\}
&= \H_j (x) \del_{x^i} \d^3(x-y) - \H_i (y) \del_{y^j} \d^3(x-y) \, . \label{covariance}
\ee
Obeying this covariance ``algebra'' is the necessary and sufficient condition for the $\H_{\mu}$ to consistently generate spacetime diffeomorphisms~\cite{Teitelboim:1972vw}.  In fact, four first class constraints obeying this algebra are guaranteed to arise in any covariant field theory; in canonical form, general covariance of the action is encoded in this constraint algebra.

For the constraints to be consistent with the equations of motion, the constraints must be preserved by the equations of motion, {\it i.e.}, $\dot{\H}_{\mu} \sim 0$.  Since $\del \H_{\mu}/\del t = 0$, applying the equations of motion to $\H_{\mu}$ yields
\be
\dot{\H}_{\mu}(x) = \int \dcy N^{\nu}(y) \{ \H_{\mu}(x), \H_{\nu}(y) \} \, .
\ee
From the first class character of the constraints, it follows that $\dot{\H}_{\mu} \sim 0$, as desired.

The Hamiltonian formulation of GR is a theory of a spatial metric $h_{ij}$ and its conjugate momentum $\pi^{i j}$, so the theory contains twelve canonical (or six real) variables.  However, these variables are not independent.  First, they are related by the four constraints $\H_{\mu} \sim 0$.  Second, the equations of motion for $h_{ij}$ and $\pi^{i j}$ depend on the four arbitrary functions $N^{\mu}$; to gauge-fix the $N^{\mu}$ would require imposing four gauge-fixing constraints.
\be
6 \cdot h_{ij} \, 's + 6 \cdot \pi^{i j} \, 's - 4 \cdot \H_{\mu} \, 's - 4 \cdot N^{\mu} \, 's = 4 {\rm \ canonical \ DoF} \, . \label{gr4dofeq}
\ee
The theory therefore has four canonical (or two real) degrees of freedom.

\section{Spatially Covariant General Relativity}
\label{gr3}
We would like to depart from general relativity by modifying the equations of motion for the two graviton degrees of freedom.  Ideally, we would like to solve all four gauge constraints, go down to the physical phase space, and modify the theory at that level.  In this way, we would circumvent all the difficulties of consistently modifying a constrained field theory.  Unfortunately, we do not know how to do this.

One possible approach is to modify the equations of motion for the phase space variables $h_{i j}$ and $\pi^{i j}$.  However, the counting of degrees of freedom in general relativity relies on the fact that the four constraints $\H_{\mu}$ satisfy a consistent first class algebra, namely the covariance algebra of equation (\ref{covariance}), and we know from the HKT theorem that any modification of the action for $h_{i j}$ and $\pi^{i j}$ will destroy this algebra.  If we modify the action for the phase space variables $h_{i j}$ and $\pi^{i j}$, we must impose an alternative constraint structure that consistently constrains the phase space to the same degree as the covariance algebra; this is the approach taken in~\cite{Horava:2010zj}.  Since we take the point of view that full spacetime covariance is a spurious symmetry, we do not wish our theory to contain a constraint structure that implies the same degree of redundancy as the covariance algebra.

Though spacetime symmetry is manifestly broken on cosmological scales (whether spontaneously or explicitly), there is strong evidence for spatial homogeneity and isotropy, so we will attempt to modify general relativity while preserving the manifest spatial covariance of the theory.  To obtain a spatially covariant formulation of general relativity to modify, we will solve the Hamiltonian constraint $\H_0$ while leaving the three momentum constraints $\H_i$ intact.  The Hamiltonian constraint is famously hard to solve in general, but we are interested in using our theories in a cosmological context, so we will solve it using a gauge-fixing constraint which is well-defined on an expanding FRW background.

\subsection{Metric Decomposition}
\label{metdec}
Before gauge-fixing, we decompose the metric $h_{i j}$ into a conformal factor $\W \equiv h^{1/3}$ and a unit-determinant metric $\th_{i j}$, {\it i.e.},
\be
h_{i j} = \W \th_{i j} \, .
\ee
Note that $\W = (\sqrt{h})^{2/3}$ is a three-scalar density of weight $2/3$, while $\th_{i j}$ is a three-tensor density of weight $-2/3$.  The scalar density we will work with is not the conformal factor $\W$, but the volume factor $\w \equiv \sqrt{h} = \W^{3/2}$, which is a scalar density of unit weight.  We choose $\w$ because its conjugate momentum,
\be
\pw \equiv \frac{\d L}{\d \dot{\w}} = \frac{2 \pi^i_{\ i}}{3\w} = - \frac{4}{3} K \, ,
\ee
is a three-scalar and hence invariant under a rescaling of $\W$ or $\w$; this fact will simplify matters in sections~\ref{ultralocal} and~\ref{realistic}.  The momentum conjugate to $\th_{i j}$ is
\be
\pt^{i j} \equiv \frac{\d L}{\d \dot{\th}_{ij}} = \W \(\pi^{i j} -  \frac{1}{3} h^{i j} \pi^k_{\ k} \)  = \w \W \( K^{i j} -  \frac{1}{3} K h^{i j} \),
\ee
which is a traceless three-tensor density of weight $5/3$; the quantity
\be
\pt^{i j}_T \equiv \frac{\pt^{i j}}{\w \W}
\ee
is the corresponding traceless three-tensor.  By defining the traceless projection tensor $\td_{i j}^{k l}$
\be
\td_{i j}^{k l} &\equiv \d_{i j}^{k l} -  \frac{1}{3} \th_{i j} \th^{k l} \, , \nn \\
 &= \d_{i j}^{k l} -  \frac{1}{3} h_{i j} h^{k l} \, ,
\ee
we can write $\pt^{i j}$ more compactly as
\be
\pt^{i j} = \W \td^{i j}_{k l} \pi^{k l}  = \W \w \td^{i j}_{k l} K^{k l} \, .
\ee
The phase space variables $h_{i j}$ and $\pi^{i j}$ can thus be written as
\be
h_{i j} = \w^{2/3} \th_{i j}, \qquad \pi^{i j} = \w^{-2/3} \pt^{i j} + \frac{1}{2} \th^{i j} \w^{1/3} \pw \, .
\ee
The decomposition of the spatial metric into a volume factor and a unit-determinant metric is completely general.  Though the corresponding conjugate momenta were derived by taking variational derivatives of the Einstein-Hilbert Lagrangian, the decomposition of the momentum tensor into its trace part and its traceless part is likewise completely general.  Those familiar with the techniques of numerical relativity may be reminded of the York-Lichnerowicz conformal decomposition or the BSSNOK (Baumgarte, 
Shapiro, Shibata, Nakamura, Oohara, and Kojima) formalism~\cite{Alcubierre:2008}.

\subsection{Cosmological gauge}
\label{cosmogauge}
To solve the constraint $\H_0$, we must first gauge-fix the lapse $N$ with a gauge-fixing constraint $\chi$ for which $\{ \H_0, \chi \} \nsim 0$; this renders $\H_0$ second class, and hence solvable.  Since we wish to retain explicit spatial covariance, our constraints $\H_i$ must remain first class.

In a cosmological context, it is natural to use the volume factor of the spatial metric as a clock, so that $t = t(\w)$; we call this cosmological gauge.  As mentioned in the introduction, cosmological gauge is only valid when the determinant of the spatial metric evolves monotonically, so this procedure is only valid when considering perturbative corrections to FRW spacetime.  When the evolution of $\w$ is monotonic, $t(\w)$ is an invertible function, so this gauge is equivalent to taking the volume factor $\w$ to be a function of time, {\it i.e.}, $\w = \w(t)$.

To impose cosmological gauge, we add to the canonical action of general relativity a gauge-fixing constraint
\be
\chi \equiv \w - \w(t) \, ,
\ee
along with a corresponding Lagrange multiplier $\l$.  The new gauge-fixed action is
\be
S' = \int \dt \dcx \( \pi^{i j}\dot{h}_{i j} - N^{\mu} \H_{\mu} - \l \chi \).
\ee
Varying the action with respect to $\l$ then reproduces the constraint
\be
\chi \sim 0 \, .
\ee
By direct calculation --- see appendix B for details --- one can verify that
\be
\{\H_0 (x), \chi(y)\} = \frac{1}{2} \pi^i_{\ i}(x) \d^3 (x-y) \, ;
\ee
the constraints $\H_0$ and $\chi$ are thus second class, so we expect to be able to solve them.  The only wrinkle is that
\be
\{\H_i (x), \chi (y)\} = \sqrt{h(x)} \del_{x^i} \d^3 (x-y) \, ,
\ee
so the constraints $\H_i$ are also second class!  By shuffling our constraints slightly, we can obtain a set of two second-class constraints and three first class constraints, and thereby preserve explicit spatial covariance. Indeed, since
\be
\left\{ 2 \sqrt{h(x)} \del_{x^i} \(\frac{\H_0(x)}{\pi^k_{\ k}(x)}\), \chi(y) \right\} \sim \sqrt{h(x)} \del_{x^i} \d^3 (x-y) \, ,
\ee
it follows that the combination
\be
\Ht_i &\equiv \H_i - 2 \sqrt{h} \, \del_{i} \( \frac{\H_0}{\pi^k_{\ k}} \) \nn \\
&= \H_i - 2 \sqrt{h} \, \grad_{i} \( \frac{\H_0}{\pi^k_{\ k}} \)
\ee
obeys
\be
\left\{ \Ht_i(x) , \Ht_j(y) \right\} \, \sim \,\left\{ \Ht_i(x) , \H_0(y) \right\} \, \sim \, \left\{ \Ht_i(x) , \chi(y) \right\} \, \sim \, 0 \, .
\ee
The interpretation of this result is simple.  The $\H_i$'s generate spatial diffeomorphisms, while $\H_0$ generates time translation.  A generic spatial diffeomorphism will alter the conformal factor of the spatial metric.  If the conformal factor is taken to be the measure of time, then the $\H_i$'s, by altering the conformal factor, will generate time translation, while $\H_0$, by generating time translation, will alter the conformal factor.  The $\Ht_i$'s generate spatial diffeomorphisms that preserve the conformal factor, so they must differ from the $\H_i$'s by the gradient of a compensating time translation term.

From the definition of $\Ht_i$, it is apparent that demanding $\chi \sim 0$ and $\H_{\mu} \sim 0$ is equivalent to demanding $\chi \sim 0$, $\H_0 \sim 0$, and $\Ht_i \sim 0$.  The latter set of constraints has the virtue that the $\Ht_i$ are first class, and can thus consistently represent symmetries.  We therefore take our five constraints to be the two second class constraints $\chi$ and $\H_0$ and the three first class constraints $\Ht_i$.  Using $\H_i = \Ht_i + 2 \sqrt{h} \grad_{i} \( \H_0/\pi^k_{\ k} \)$, the gauge-fixed action can be rewritten in terms of $\Ht_i$ as
\be
S' = \int \dt \dcx \( \pi^{i j}\dot{h}_{i j} - N^0 \H_0 - N^i \Ht_i - 2 \sqrt{h} N^i \grad_{i} \( \frac{\H_0}{\pi^k_{\ k}} \) - \l \chi \).
\ee
Upon integration by parts, the action becomes
\be
S' = \int \dt \dcx \( \pi^{i j}\dot{h}_{i j} - \Nt \H_0 - N^i \Ht_i - \l \chi \),
\ee
where
\be
\Nt \equiv N - 2 \frac{\sqrt{h}}{\pi^k_{\ k}} \grad_{i} N^i \, .
\ee
Variation of the action $S'$ with respect to $h_{i j}$ and $\pi^{i j}$ yields Hamilton's equations,
\be
\dot{h}_{i j}(x) = \frac{\d H'}{\d \pi^{i j}(x)} \, , \qquad \dot{\pi}^{i j}(x) = -\frac{\d H'}{\d h_{i j}(x)} \, ,
\ee
where the new Hamiltonian $H'$ is
\be
H' = \int \dcx \( \Nt \H_0 + N^i \Ht_i + \l \chi \).
\ee
The equation of motion for any quantity $A(h_{i j}, \pi^{i j}, t)$ is therefore
\be
\dot{A}
&= \frac{\del A}{\del t} + \{ A, H' \} \nn \\
&= \frac{\del A}{\del t} + \int \dcy \( \Nt(y) \{ A, \H_0(y) \} + N^i(y) \{ A, \Ht_i(y) \} + \l(y) \{ A, \chi(y) \} \) ,
\ee
where the Poisson bracket is defined as in (\ref{pbgr}).  Variation of the action $S'$ with respect to $\Nt$, $\l$, and $N^i$ yields the five constraints
\be
\H_0 \sim 0 \, , \qquad \chi \sim 0 \, , \qquad \Ht_i \sim 0 \, .
\ee
The action does not contain time derivatives of the Lagrange multipliers, so at first their evolution appears unconstrained.  Since the $\Ht_i$ are first class, the three functions $N^i$ are indeed arbitrary until and unless we gauge-fix them.  The evolution of $\Nt$ and $\l$, however, will be determined by demanding the consistency of $\H_0$ and $\chi$ with the equations of motion.

For the constraints to be consistent with the equations of motion, they must be preserved by the equations of motion; we therefore demand that $\dot{\Ht}_i \sim 0$, $\dot{\H}_0 \sim 0$, and $\dot{\chi} \sim 0$.  Since the $\Ht_i$ are first class and $\del_t \Ht_i = 0$, it follows at once that $\dot{\Ht}_i \sim 0$.  Since $\del_t \H_0 = 0$, $\{\H_0(x), \H_0(y) \} \sim 0$, and $\{\H_0(x), \Ht_i(y) \} \sim 0$, it follows that
\be
\dot{\H}_0(x) &\sim \int \dcy \l(y) \{ \H_0(x), \chi(y) \} \nn \\
&\sim \frac{1}{2} \l(x) \pi^i_{\ i}(x) \, .
\ee
On a flat FRW background,\footnote{A spatially-flat FRW spacetime corresponds to $N = 1$, $N_i = 0$, and $h_{i j} = a^2(t) \d_{i j}$.} $K = 3 \dot{a}/a$ and hence $\pi^i_{\ i} = -2 \w K = -6 \dot{a} a^2$.  Since we are only considering gravity on an expanding background, we assume that $\pi^i_{\ i}(x) \nsim 0$ more generally.  The demand $\dot{\H}_0 \sim 0$ thus implies
\be
\l \sim 0 \, .
\ee
Since $\{\chi(x), \chi(y) \} = 0$, $\{\chi(x), \Ht_i(y) \} \sim 0$, and $\del \chi/\del t = -\dot{\w}(t)$, it follows that
\be
\dot{\chi}(x)
&\sim -\dot{\w}(t) + \int \dcy \Nt(y) \{ \chi(x), \H_0(y) \}, \nn \\
&\sim -\dot{\w}(t) -\frac{1}{2} \Nt(x) \pi^i_{\ i}(x) \, .
\ee
Since $\pi^i_{\ i} \nsim 0$, demanding $\dot{\chi} \sim 0$ allows us to solve for $\Nt$,
\be
\Nt \sim \frac{-2 \dot{\w}(t)}{\pi^i_{\ i}} \, .
\ee
The functions $\Nt$ and $\l$ are thus not arbitrary.  Since $N = \Nt + 2 \sqrt{h} (\grad_{i} N^i)/\pi^k_{\ k}$, the lapse $N$ has not been completely gauge-fixed, but its arbitrariness stems solely from its dependence on the three arbitrary functions $N^i$.

As a check, let us revisit the counting of degrees of freedom in cosmological gauge.  For these purposes, the only effect of gauge-fixing is to replace the first class constraint $\H_0 \sim 0$ and the arbitrary function $N$ with the second class constraints $\H_0 \sim 0$ and $\chi \sim 0$.  This modifies the left hand side of equation~(\ref{gr4dofeq}), but does not change the final tally.
\be
6 \cdot h_{ij} \, 's + 6 \cdot \pi^{i j} \, 's - 1 \cdot \H_0 - 1 \cdot \chi - 3 \cdot \Ht_i \, 's - 3 \cdot N^i \, 's = 4 {\rm \ canonical \ DoF} \, . \label{cosmogaugedofeq}
\ee
After gauge-fixing, the theory still has four canonical (or two real) degrees of freedom.

\subsection{Solving $\H_0$ and $\chi$}
\label{solveH0}
In this section, we will solve the constraints $\H_0$ and $\chi$ to obtain a spatially covariant formulation of general relativity as a theory of a unit-determinant metric $\th_{i j}$ and its conjugate momentum $\pt^{i j}$.  This will set the stage for modifying general relativity in section~\ref{ultralocal}.

Since $\chi$ and $\H_0$ are second class, they can be solved explicitly to yield expressions for $\w$ and $\pw$ in terms of $t$, $\th_{i j}$, $\pt^{i j}$, and spatial derivatives.  ``Solving'' for $\w$ is trivial: $\w = \w(t)$.  Solving for $\pw$ requires us to take a square root and pick a sign, which amounts to picking either an expanding or a contracting background.  We pause to emphasize once again that our procedure is only valid in a cosmological context, when the conformal factor of the spatial metric can be assumed to be evolving monotonically.  To pick the sign corresponding to an expanding background, first recall that
\be
\pw = -\frac{4}{3} K \, .
\ee
On a flat FRW background, $K = 3 \dot{a}/a$ and hence $\pw = - 4 \dot{a}/a$.  An expanding FRW background therefore corresponds to $\pw < 0$.  Returning to the general case, we choose $\pw<0$ to obtain
\be
\pw = \pgr \equiv -\sqrt{\frac{8}{3}} \, \sqrt{ \frac{\pt_{i j} \pt^{i j}}{\w^2} - \frac{\Rt}{\w^{2/3}} + 2 \L} \, , \label{pigr}
\ee
where indices are raised and lowered with $\th_{i j}$, and $\Rt$ is the Ricci scalar for $\th_{i j}$.  Substituting these results for $\w$ and $\pw$ back into the action $S'$ yields the action of general relativity on the reduced phase space ($\th_{i j}$, $\pt^{i j}$),
\be
S'' = \int \dt \dcx \( \, \pt^{i j} \dot{\th}_{i j} + \pw \dot{\w} - N^i \Ht_i \, \),
\ee
where
\be
\Ht_i = -2 \th_{i j} \gradt_k \pt^{j k} - \w \gradt_i \pw \, ,
\ee
and $\gradt_i$ is the covariant derivative with respect to $\th_{i j}$.  This action yields the new Hamiltonian
\be
H'' = \int \dcx \( \, - \dot{\w} \pw + N^i \Ht_i \, \) \, .
\ee
The term $\pi^{i j} \dot{h}_{i j}$ has split into the term $\pt^{i j} \dot{\th}_{i j}$ and a contribution $-\dot{\w} \pw$ to the physical Hamiltonian density.  Variation of the action with respect to $\th_{i j}$ and $\pt^{i j}$ yields
\be
\dot{\th}_{i j}(x) = \frac{\d H''}{\d \pt^{i j}(x)} \, , \qquad \dot{\pt}^{i j}(x) = -\frac{\d H''}{\d \th_{i j}(x)} \, .
\ee
To evaluate these variational derivatives, one must use the relations
\be
\frac{\d \th_{i j}(x)}{\d \th_{k l}(y)} = \td^{k l}_{i j} \d^3(x-y) \, , \qquad \frac{\d \pt^{i j}(x)}{\d \th_{k l}(y)} = -\frac{1}{3} \th^{i j} \pt^{k l} \d^3(x-y) \, ,
\ee
and
\be
\frac{\d \th_{i j}(x)}{\d \pt^{k l}(y)} = 0 \, , \qquad \frac{\d \pt^{i j}(x)}{\d \pt^{k l}(y)} = \td^{i j}_{k l} \d^3(x-y) \, .
\ee
Defining the Poisson bracket appropriate to the reduced phase space,
\be
\{ A, B \} \equiv \int \dcx \( \frac{\d A}{\d \th_{i j}(x)} \frac{\d B}{\d \pt^{i j}(x)} - \frac{\d A}{\d \pt^{i j}(x)} \frac{\d B}{\d \th_{i j}(x)}\) \, , \label{reducedPB}
\ee
any quantity $A(\th_{i j}, \pt^{i j}, t)$ obeys the equation of motion
\be
\dot{A} = \frac{\del A}{\del t} + \{ A, H'' \} \, .
\ee
Variation of the action with respect to $N^i$ yields the three constraints
\be
\Ht_i \sim 0 \, .
\ee
As before, the time evolution of $N^i$ is unconstrained by the action; in the absence of a gauge-fixing procedure, the three functions $N^i$ are arbitrary.

\subsection{Constraint Properties \& Degrees of Freedom}
\label{gr3dof}
By lengthy direct calculation, it is possible to prove that the constraints $\Ht_i$ are first class, {\it i.e.}, $\{\Ht_i(x),\Ht_j(y)\} \sim 0$.  Furthermore, by applying the equations of motion to $\Ht_i$, it is possible to show that $\dot{\Ht}_{i} \sim 0$, so the constraints are preserved by the equations of motion.  We defer demonstrations of these two facts to section~\ref{realistic}, where we will examine general relativity in the context of a class of realistic theories.  This is an important consistency check, because {\it a priori} it is not clear that our procedure for solving the Hamiltonian constraint will yield a consistent action on the reduced phase space.

As a final check, we revisit the counting of degrees of freedom in spatially covariant general relativity.  After imposing cosmological gauge and solving the Hamiltonian constraint $\H_0 \sim 0$, general relativity is a theory of a unit-determinant spatial metric $\th_{ij}$ and its traceless conjugate momentum $\pt^{i j}$, so the theory contains ten canonical (or 5 real) variables.  This reduction in the size of the phase space is compensated by a corresponding reduction in the number of constraints and arbitrary functions: the theory contains three first class constraints $\Ht_i \sim 0$, and its equations of motion involve three arbitrary functions $N^i$.
\be
5 \cdot \th_{ij} \, 's + 5 \cdot \pt^{i j} \, 's - 3 \cdot \Ht_i \, 's - 3 \cdot N^i \, 's = 4 {\rm \ canonical \ DoF} \, .
\ee
Spatially covariant general relativity thus contains four canonical (or two real) degrees of freedom, the same number as fully covariant general relativity.

\section{Ultralocal Modified Gravity}
\label{ultralocal}
We have two criteria in mind for our modified theories of gravity: two graviton degrees of freedom, and manifest spatial covariance.  Our starting point is the action of spatially covariant general relativity, which has both of these properties.  To modify general relativity, we will change the functional form of the scalar quantity $\pw$, which in general relativity obeys $\pw = \pgr$.  This yields the action
\be
S = &\int \dt \dcx \( \, \pt^{i j} \dot{\th}_{i j} + \dot{\w} \pw - N^i \Ht_i \, \) , \nn \\
\Ht_i &= -2 \th_{i j} \gradt_k \pt^{j k} - \w \gradt_i \pw \, ,
\ee
where $\pw$ is an unspecified scalar function of $t$, the phase space variables $\th_{i j}$ and $\pt^{i j}$, and spatial derivatives.  This action leads to the equation of motion
\be
\dot{A} = \frac{\del A}{\del t} + \{ A, H \} \, , \label{eom}
\ee
where the Hamiltonian $H$ is
\be
H = \int \dcx \( \, - \dot{\w} \pw + N^i \Ht_i \, \) ,
\ee
and the Poisson bracket is
\be
\{ A, B \} \equiv \int \dcx \( \frac{\d A}{\d \th_{i j}(x)} \frac{\d B}{\d \pt^{i j}(x)} - \frac{\d A}{\d \pt^{i j}(x)} \frac{\d B}{\d \th_{i j}(x)}\) \, . \label{pbmg}
\ee
Retaining the manifest spatial covariance of the theory amounts to demanding 1) that the modified $\Ht_i$ remain first class, {\it i.e.},
\be
\{\Ht_i(x),\Ht_j(y)\} \sim 0 \, ,
\ee
and 2) that the modified constraints be preserved by the modified equations of motion, {\it i.e.},
\be
\dot{\Ht}_i \sim 0 \, .
\ee
Any theory satisfying these two points will be manifestly covariant under spatial diffeomorphisms, with the constraints $\Ht_i$ acting as the generators of the gauge symmetry.  Moreover, the presence of three first class constraints $\Ht_i$ on the phase space $(\th_{i j}, \pt^{i j})$ guarantees that such a theory contains two local degrees of freedom, exactly as desired.

In the remainder of the paper, we examine two classes of theories.  First, for pedagogical purposes, we treat the case when $\pw$ is an ultralocal\footnote{Local functions depend on the value of fields in a neighborhood around a point, so they can be written in terms of fields and derivatives of fields.  Ultralocal functions only depend on the value of fields {\it at} a point, so they do not contain derivatives.  The ultralocal limit is commonly used to analyze long distance cosmological perturbations.  The idea of using the cosmological gauge in the ultralocal limit, sometimes referred to as the separate universes approach, is treated in~\cite{Tolley:2008na}.} function of time $t$ and the phase space variables $\th_{i j}$ and $\pt^{i j}$; in other words, $\pw$ will not contain spatial derivatives.  Second, to make contact with general relativity, we treat the more realistic case when $\pw$ also depends on $\Rt$, the Ricci scalar of $\th_{i j}$.  Forthcoming work will examine more general classes of scalar momenta~\cite{godfreytoappear}.  In this section, we use the ultralocal case to introduce the formalism needed to determine when the constraints $\Ht_i$ remain first class and when the constraints are preserved by the equations of motion.  In section~\ref{realistic}, we apply the formalism to the realistic case.  In both the ultralocal and the realistic case, the consistency of the constraints with the equations of motion requires $\pw$ to satisfy an analogue of the renormalization group equation; scalar momenta satisfying this equation are manifestly invariant under rescaling of the volume factor $\w$.  In the ultralocal case, this is the only consistency condition that arises.  In the realistic case, demanding that the constraints $\Ht_i$ satisfy a first class algebra is equivalent to demanding that $\pw$ obey a rather complicated differential equation.

\subsection{Constraint Algebra}
\label{ulca}
In this section, we will compute the Poisson bracket $\{ \Ht_i(x), \Ht_a(y) \}$ assuming that $\pw$ is an ultralocal function, and use the result to determine when the constraints $\Ht_i$ remain first class.  To simplify the calculation of $\{ \Ht_i(x), \Ht_a(y) \}$, we split $\Ht_i$ into a tensor part $\J_i$ and a scalar part $\K_i$.  Concretely, we define the vector densities
\be
\J_i \equiv - 2 \th_{i j} \gradt_k \pt^{j k} \, , \qquad \K_i \equiv - \w \gradt_i \pw \, ,
\ee
in terms of which $\Ht_i$ becomes simply
\be
\Ht_i = \J_i + \K_i \, .
\ee
The Poisson bracket $\{ \Ht_i(x), \Ht_a(y) \}$ can then be written as the sum of more manageable brackets,
\be
\{ \Ht_i(x), \Ht_a(y) \}
&= \{ \J_i(x), \J_a(y) \} + \{ \K_i(x), \K_a(y) \} \nn \\
&+ \{ \J_i(x), \K_a(y) \} + \{ \K_i(x), \J_a(y) \} \, . \label{hjk}
\ee
To simplify the evaluation of these component Poisson brackets, we introduce the smoothing functionals
\be
F_J \equiv \int \dcx f^i \J_i \, , &\qquad F_K \equiv \int \dcx f^i \K_i \, , \nn \\
G_J \equiv \int \dcy g^a \J_a \, , &\qquad G_K \equiv \int \dcy g^a \K_a \, , \label{fjfk}
\ee
where the functions $f^i$ and $g^i$ are time-independent smoothing functions.  We then compute the brackets
\be
\{ F_J, G_J \} &= \int \dcx \dcy f^i(x) g^a(y) \{ \J_i(x),\J_a(y) \} \, , \nn \\
\{ F_J, G_K \} + \{ F_K, G_J \} &= \int \dcx \dcy f^i(x) g^a(y) \bigg( \{ \J_i(x),\K_a(y) \} + \{ \K_i(x), \J_a(y) \} \bigg) , \nn \\
\{ F_K, G_K \} &= \int \dcx \dcy f^i(x) g^a(y) \{ \K_i(x),\K_a(y) \} \, . \label{intjk}
\ee
We make the key assumption that the smoothing functions decay so rapidly at infinity that when we integrate by parts inside the smoothing functionals, the boundary term vanishes identically; the smoothing functions are otherwise arbitrary.  With the freedom to integrate by parts at will, it is straightforward to compute variational derivatives of the smoothing functionals, and thereby to obtain explicit expressions for their Poisson brackets.  By comparing these explicit expressions to the formal expressions in equation~(\ref{intjk}), we will derive explicit expressions for the Poisson brackets involving $\J_i$ and $\K_i$.

To compute variational derivatives of the smoothing functional $F_J$, first integrate by parts to obtain
\be
F_J = 2 \int \dcx \th_{i j} \pt^{j k} \gradt_k f^i \, ,
\ee
from which it follows that
\be
\d F_J = \int \dcx \left\{ 2 \pt^{j k} (\gradt_k f^i) \d \th_{i j} + 2 \th_{i j} (\gradt_k f^i) \d \pt^{j k} + 2 \th_{i j} \pt^{j k} \d \gradt_k f^i \right\}. \label{dfjinit}
\ee
The first two terms in this integral are in a convenient form for taking variational derivatives with respect to $\th_{i j}$ and $\pt^{j k}$, but the third term requires finessing.  To evaluate $\d \gradt_{k} f^i$, expand the covariant derivative as $\gradt_{k} f^i = \del_{k} f^i + \Gt^{i}_{k r} f^r$, where $\Gt_{j k}^i$ is the connection of the metric $\th_{i j}$.  It follows immediately that $\d \gradt_{k} f^i = f^r \d \Gt^{i}_{k r}$.  The identity
\be
\d \Gt^{i}_{k r} = \frac{1}{2} \th^{i m} \( \grad_r \d \th_{k m} + \grad_k \d \th_{r m} - \grad_m \d \th_{r k} \)
\ee
thus implies that $ 2 \pt^{j k} \th_{i j} \d \gradt_{k} f^i = f^i \pt^{j k} \gradt_i \d \th_{j k}$, so equation~(\ref{dfjinit}) becomes
\be
\d F_J = \int \dcx \left\{ 2 \pt^{j k} (\gradt_k f^i) \d \th_{i j} + 2 \th_{i j} (\gradt_k f^i) \d \pt^{j k} + f^i \pt^{j k} \gradt_i \d \th_{j k} \right\}.
\ee
Integrating by parts, this reduces to
\be
\d F_J = \int \dcx \left\{ 2 \pt^{j k} (\gradt_k f^i) \d \th_{i j} - \gradt_i (f^i \pt^{j k}) \d \th_{j k} + 2 \th_{i j} (\gradt_k f^i) \d \pt^{j k} \right\}.
\ee
From this expression, it is straightforward to compute variational derivatives of $F_J$,
\be
\frac{\d F_J}{\d \th_{m n}} &= 2 \, \td^{m n}_{i j} \, \pt^{j k} \gradt_k f^i - \gradt_i \( f^i \pt^{m n} \) - \frac{2}{3} \pt^{m n} \gradt_i f^i \, , \nn \\
\frac{\d F_J}{\d \pt^{m n}} &= 2 \, \td^{j k}_{m n} \, \th_{i j} \gradt_k f^i \, . \label{fjhp}
\ee
The corresponding results for $G_J$ are
\be
\frac{\d G_J}{\d \th_{m n}} &= 2 \, \td^{m n}_{a b} \, \pt^{b c} \gradt_c g^a - \gradt_a \( g^a \pt^{m n} \) - \frac{2}{3} \pt^{m n} \gradt_a g^a \, , \nn \\
\frac{\d G_J}{\d \pt^{m n}} &= 2 \, \td^{b c}_{m n} \, \th_{a b} \gradt_c g^a \, . \label{gjhp}
\ee

The variational calculation for the smoothing functional $F_K$ is less straightforward.  After integrating by parts, $F_K$ becomes
\be
F_K = \w \int \dcx \( \del_i f^i \) \pw \, ,
\ee
from which it follows that
\be
\d F_{K} = \w \int \dcx \( \del_i f^i \) \d \pw \, . \label{varfk}
\ee
To evaluate $\d \pw$ in full generality would be very difficult, so we will make some simplifying assumptions about the form of $\pw$.  In this section, we will assume that $\pw$ is an ultralocal function of $t$, $\th_{i j}$, and $\pt^{i j}$.

To facilitate calculations, we will enumerate all the scalars that can be built by contracting factors of $\th_{i j}$ against factors of $\pt^{i j}$.  We begin by recursively defining $\Pi^{i j}(n)$, the linked chain of $n$ factors of $\pt^{i j}$.  The chain of zero factors of $\pt^{i j}$ is simply
\be
\Pi^{i j}(0) \equiv \th^{i j} \, .
\ee
The process of adding a link to the chain is defined by
\be
\Pi^{i j}(n+1) \equiv \pt^{i}_{\ k} \Pi^{k j}(n) \, .
\ee
By closing the chain, one obtains scalars,
\be
\phi(n) \equiv \Pi^{i}_{\ i}(n) \, . \label{phin}
\ee
The $\phi(n)$ are the only scalars that can be built out of connected contractions of $\th_{i j}$ and $\pt^{i j}$.  For an arbitrary ultralocal function $\pw$, it follows that
\be
\d \pw = \sum_{n = 2}^{\infty} \frac{\del \pw}{\del \phi(n)} \d \phi(n) \, .
\ee
Since $\phi(0) = 3$ and $\phi(1) = 0$, $\d \phi(0) = \d \phi(1) = 0$.  For $n \geq 2$, the variational derivatives of the $\phi(n)$ are
\be
\frac{\d \phi(n) (x)}{\d \th_{m n}(y)}
&= \( \td^{m n}_{a b} n \Pi(n)^{a b} - \frac{1}{3} \pt^{m n} n \phi(n-1) \) \d^3(x-y) \, , \nn \\
\frac{\d \phi(n) (x)}{\d \pt^{m n}(y)}
&= \td^{a b}_{m n} n \Pi(n-1)_{a b} \d^3(x-y) \, . \label{deltaphidelta}
\ee
The variational derivatives of $F_K$ are thus
\be
\frac{\d F_K}{\d \th_{m n}}
&= \w \( \del_i f^i \) \sum_{n = 2}^{\infty} n \frac{\del \pw}{\del \phi(n)} \( \td^{m n}_{j k} \Pi(n)^{j k} - \frac{1}{3} \pt^{m n} \phi(n-1) \) , \nn \\
\frac{\d F_K}{\d \pt^{m n}}
&= \w \( \del_i f^i \) \sum_{n = 2}^{\infty} n \frac{\del \pw}{\del \phi(n)} \td^{j k}_{m n} \Pi(n-1)_{j k} \, . \label{fkhp}
\ee
Similarly, the variational derivatives of $G_K$ are
\be
\frac{\d G_K}{\d \th_{m n}}
&= \w \( \del_a g^a \) \sum_{m = 2}^{\infty} m \frac{\del \pw}{\del \phi(m)} \( \td^{m n}_{b c} \Pi(m)^{b c} - \frac{1}{3} \pt^{m n} \phi(m-1) \) , \nn \\
\frac{\d G_K}{\d \pt^{m n}}
&= \w \( \del_a g^a \) \sum_{m = 2}^{\infty} m \frac{\del \pw}{\del \phi(m)} \td^{b c}_{m n} \Pi(m-1)_{b c} \, . \label{gkhp}
\ee
We emphasize that these results for $F_K$ and $G_K$ rely on the ultralocality assumption, and will be modified in section~\ref{realistic}.

We are now in a position to compute the Poisson brackets of the smoothing functionals, from which we will extract the Poisson brackets of the vector densities $\J_i$ and $\K_i$.

\begin{itemize}
\item $\{ \J_{i}(x), \J_{a}(y) \}$

To obtain the bracket $\{ \J_{i}(x), \J_{a}(y) \}$, we first compute $\{ F_J, G_J \}$.  Combining the $F_J$ and $G_J$ variations into the bracket $\{ F_J, G_J \}$ yields
\be
\{ F_J, G_J \} = 2 &\int \dcz \left\{ \(\gradt_c f^i \) \(\gradt_i g^a \) \th_{a b} \pt^{b c} - \(\gradt_k g^a \) \(\gradt_a f^i \) \th_{i j} \pt^{j k} \right. \nn \\
 &\left.+ \(\gradt_k f^i\) \gradt_a \( g^a \th_{i j} \pt^{j k} \) - \(\gradt_c g^a\) \gradt_i \( f^i \th_{a b} \pt^{b c} \) \right\} .
\ee
After integrating by parts, using the definition $\J_i = - 2 \th_{i j} \gradt_k \pt^{j k}$, and using the identity $\(\gradt_i \gradt_j - \gradt_j \gradt_i \) V^a = \Rt^a_{\ b i j} V^b$, this reduces to
\be
\{ F_J, G_J \} = &\int \dcz \left\{ f^i \J_a \gradt_i g^a - g^a \J_i \gradt_a f^i + 2 f^i g^a \pt^{j k} \( \Rt_{j i k a} + \Rt_{j a i k} \) \right\} .
\ee
From the symmetries of the Riemann tensor\footnote{$\Rt_{a b c d} = \Rt_{c d a b}$, $\Rt_{a b c d} = - \Rt_{b a c d} = - \Rt_{a b d c}$.} and the traceless momentum tensor\footnote{$\pt^{i j} = \pt^{j i}$.}, it follows that $0 = \pt^{j k} \( \Rt_{j i k a} + \Rt_{j a i k} \)$, so the last term in the integrand vanishes.  The connection terms inside the remaining covariant derivatives cancel to yield
\be
\{ F_J, G_J \} = \int \dcz \( f^i \J_a \del_i g^a - g^a \J_i \del_a f^i \) .
\ee
To extract the bracket $\{ \J_{i}(x), \J_{a}(y) \}$ from this result, first relabel dummy indices
\be
\{ F_J, G_J \} = \int \dcx f^i \J_a \del_i g^a - \int \dcy g^a \J_i \del_a f^i \, .
\ee
Under the spatial derivatives in this equation, insert the identities
\be
g^a(x) = \int \dcy \d^3(x-y) g^a(y) \, , \qquad f^i(y) = \int \dcx \d^3(x-y) f^i(x) \, , \label{deltaeq}
\ee
to obtain
\be
\{ F_J, G_J \} = \int \dcx \dcy f^i(x) g^a(y) \( \J_a(x) \del_{x^i} \d^3(x-y) - \J_i(y) \del_{y^a} \d^3(x-y) \) .
\ee
Comparing this expression to equation~(\ref{intjk}) yields the identity
\be
\{ \J_i(x),\J_a(y) \} = \J_a(x) \del_{x^i} \d^3(x-y) - \J_i(y) \del_{y^a} \d^3(x-y) \, . \label{jj}
\ee
This is the same algebra obeyed by the $\H_i$ in equation~(\ref{covariance}).  This result is completely independent of our choice of $\pw$, and will carry over unchanged into section~\ref{realistic}.

\item  $\{ \J_i(x), \K_a(y) \}+\{ \K_i(x), \J_a(y) \}$

To obtain $\{ \J_i(x), \K_a(y) \}+\{ \K_i(x), \J_a(y) \}$, we first compute $\{ F_J, G_K \} + \{ F_K, G_J \}$.  Assembling the $F_J$ and $G_K$ variations into the Poisson bracket $\{ F_J , G_K \}$ yields
\be
\{ F_J , G_K \} = - \w \int \dcz \( \del_a g^a \) \sum_{m = 2}^{\infty} \frac{\del \pw}{\del \phi(m)} m \Pi(m-1)_{b c} \gradt_i \( f^i \pt^{b c} \) .
\ee
By expanding the covariant derivative, simplifying the ensuing total derivative of $\pw$, and recalling that $\K_i = - \w \gradt_i \pw$, this expression reduces to
\be
\{ F_J , G_K \} = \int \dcz f^i \K_i \del_a g^a - \w \int \dcz \(\del_i f^i \) \( \del_a g^a \) \sum_{m = 2}^{\infty} \frac{\del \pw}{\del \phi(m)} m \phi(m) \, . \label{fjgkforpw}
\ee
Similarly,
\be
\{ F_K, G_J \} = - \int \dcz g^a \K_a \del_i f^i + \w \int \dcz \(\del_i f^i \) \( \del_a g^a \) \sum_{m = 2}^{\infty} \frac{\del \pw}{\del \phi(m)} m \phi(m) \, ,
\ee
so the sum of the two brackets simplifies considerably,
\be
\{ F_J , G_K \} + \{ F_K , G_J \} = \int \dcz \( f^i \K_i \del_a g^a - g^a \K_a \del_i f^i \) .
\ee
Integrating by parts and invoking the identity $\del_i \K_a = \del_a \K_i$ yields
\be
\{ F_J , G_K \} + \{ F_K , G_J \} = &\int \dcz \( f^i \K_a \del_i g^a - g^a \K_i \del_a f^i \) .
\ee
To extract the quantity $\{ \J_i(x), \K_a(y) \}+\{ \K_i(x), \J_a(y) \}$, relabel dummy indices and insert the identities in equation~(\ref{deltaeq}) to obtain
\be
\{ F_J , G_K \} + \{ F_K , G_J \} = \int \dcx & \dcy f^i(x) g^a(y) \nn \\
&\times \( \K_a(x) \del_{x^i} \d^3(x-y) - \K_i(y) \del_{y^a} \d^3(x-y) \) .
\ee
Combined with equation~(\ref{intjk}), this result implies that
\be
\{ \J_i(x),\K_a(y) \} + \{ \K_i(x),\J_a(y) \} = \K_a(x) \del_{x^i} \d^3(x-y) - \K_i(y) \del_{y^a} \d^3(x-y) \, . \label{jk}
\ee
This expression depends strongly on the assumed form for $\pw$.  This result is modified heavily in section~\ref{rconstraint}, when $\pw$ is allowed to depend on $\Rt$.

\item $\{ \K_i(x), \K_a(y) \}$

To obtain $\{ \K_i(x), \K_a(y) \}$, we first compute the bracket $\{ F_K, G_K \}$.  Substituting the $F_K$ and $G_K$ variations into the Poisson bracket $\{ F_K , G_K \}$ yields
\be
\{ F_K , G_K \} = \w^2 \( \del_i f^i \) \( \del_a g^a \) &\sum_{m = 2}^{\infty} \sum_{n = 2}^{\infty} m n \frac{\del \pw}{\del \phi(m)} \frac{\del \pw}{\del \phi(n)} \nn \\
&\times \( \Pi(n)^{b c} \Pi(m-1)_{b c} - \Pi(m)^{j k}\Pi(n-1)_{j k} \) .
\ee
From the definition of the momentum chain $\Pi(n)^{i j}$, it follows that $\Pi(n)^{b c} \Pi(m-1)_{b c} = \Pi(m)^{j k}\Pi(n-1)_{j k} = \phi(n+m-1)$.  The terms of the sum thus vanish order by order, so the bracket reduces to
\be
\{ F_K , G_K \} = 0 \, . \label{fkgkforpw}
\ee
By comparing this result to equation~(\ref{intjk}), it is apparent that
\be
\{ \K_i(x), \K_a(y) \} = 0 \, . \label{kk}
\ee
When $\pw$ is an ultralocal function of the phase space variables, the Poisson bracket $\{ \K_i(x), \K_a(y) \}$ vanishes identically.  This will not be the case when $\pw$ depends nontrivially on $\Rt$, as in section~\ref{rconstraint}.
\end{itemize}
By substituting equations~(\ref{jj}),~(\ref{jk}), and~(\ref{kk}) into equation~(\ref{hjk}), and recalling that $\Ht_i = \J_i + \K_i$, we obtain
\be
\{ \Ht_i(x) , \Ht_j(y) \} = \Ht_j(x) \del_{x^i} \d^3(x-y) - \Ht_i(y) \del_{y^j} \d^3(x-y) \, .
\ee
This is the same algebra obeyed by the $\H_i$ in equation~(\ref{covariance}), and by the $\J_i$ in equation~(\ref{jj}).  Since $\Ht_i \sim 0$, this result implies that $\{ \Ht_i(x) , \Ht_j(y) \} \sim 0$, so the constraints $\Ht_i$ are first class.  To establish this result, we assumed only that $\pw$ was an arbitrary ultralocal function of $t$, $\th_{i j}$, and $\pt^{i j}$; we showed that this was equivalent to making $\pw$ a function of $t$ and the scalars $\phi(n)$ defined in equation~(\ref{phin}).  Evidently, $\pw$ can be made any ultralocal function of the phase space variables and the momentum constraints will remain first class.

\subsection{Consistency of Constraints with Equations of Motion}
\label{ulconsistent}
In this section, we will compute the time derivative $\dot{\Ht}_i$ assuming that $\pw$ is an ultralocal function, and use the result to determine when the constraints $\Ht_i$ are preserved by the equations of motion.  The time evolution of $\Ht_i$ is determined by the equation of motion
\be
\dot{\Ht}_i = \frac{\del \Ht_i}{\del t} + \{ \Ht_i, H \} \, , \label{dot}
\ee
where
\be
H = \int \dcx \( \, - \dot{\w} \pw + N^i \Ht_i \, \) \, .
\ee
Since $\Ht_i = \J_i + \K_i$ and $\del \J_i/\del t = 0$, it follows that $\del \Ht_i/\del t = \del \K_i/\del t$.  Recalling that $\K_i = - \w \del_i \pw$, the first term in equation~(\ref{dot}) becomes
\be
\frac{\del \Ht_i}{\del t} = - \del_i \( \dot{\w} \pw + \w \frac{\del \pw}{\del t}\) \, . \label{delhdelt}
\ee
To simplify the bracket $\{ \Ht_i, H \}$, we define
\be
\Pw \equiv \int \dcx \pw \, ,
\ee
so that $H$ can be written as
\be
H = - \dot{\w} \Pw + \int \dcx N^i \Ht_i \, .
\ee
From the first class character of the constraints $\Ht_i$, it follows that $\{ \Ht_i, H \} \sim - \dot{\w} \{ \Ht_i, \Pw \}$.  Since $\Ht_i = \J_i + \K_i$, the second term in equation~(\ref{dot}) becomes
\be
\{ \Ht_i, H \} \sim - \dot{\w} \{ \J_i, \Pw \} - \dot{\w} \{ \K_i, \Pw \} \, . \label{htiH}
\ee
To compute the brackets $\{ \J_i, \Pw \}$ and $\{ \K_i, \Pw \}$, we first compute the smoothing functional brackets
\be
\{ F_J, \Pw \} &= \int \dcx f^i(x) \{ \J_i(x), \Pw \} \nn \\
\{ F_K, \Pw \} &= \int \dcx f^i(x) \{ \K_i(x), \Pw \} \, . \label{fjkpw}
\ee
We have already done all the work needed to evaluate these two brackets: since $\Pw$ can be obtained from $G_K$ by the substitution $\del_a g^a \gt \w^{-1}$, brackets involving $\Pw$ can be obtained by applying this substitution to brackets involving $G_K$.

\begin{itemize}
\item $\{ \J_i , \Pw \}$

To compute the bracket $\{ \J_i, \Pw \}$, we first compute the bracket $\{ F_J, \Pw \}$.  Applying $\del_a g^a \gt \w^{-1}$ to equation~(\ref{fjgkforpw}) and integrating by parts yields
\be
\{ F_J , \Pw \} = \int \dcx f^i \del_i \(- \pw + \sum_{m = 2}^{\infty} m \phi(m) \frac{\del \pw}{\del \phi(m)} \) \, .
\ee
It follows by comparing this result with equation~(\ref{fjkpw}) that
\be
\{ \J_i , \Pw \} = \del_i \( - \pw + \sum_{m = 2}^{\infty} m \phi(m) \frac{\del \pw}{\del \phi(m)} \) \, . \label{jph}
\ee
\item $\{ \K_i , \Pw \}$

To compute the bracket $\{ \K_i, \Pw \}$, we first compute the bracket $\{ F_K, \Pw \}$.  By applying the transformation $\del_a g^a \gt \w^{-1}$, equation~(\ref{fkgkforpw}) becomes
\be
\{ F_K, \Pw \} = 0 \, .
\ee
Along with equation~(\ref{fjkpw}), this implies that
\be
\{ \K_i , \Pw \} = 0 \, . \label{kph}
\ee

\end{itemize}
By substituting equations~(\ref{jph}) and~(\ref{kph}) into equation~(\ref{htiH}), we obtain
\be
\{ \Ht_i, H \} \sim \dot{\w} \del_i \(\pw - \sum_{m = 2}^{\infty} m \phi(m) \frac{\del \pw}{\del \phi(m)} \) \, . \label{htiH2}
\ee
Upon inserting equations~(\ref{htiH2}) and~(\ref{delhdelt}) into the equation of motion~(\ref{dot}), the $\dot{\w} \del_i \pw$ terms cancel to yield
\be
\dot{\Ht}_i \sim - \del_i \(\w \frac{\del \pw}{\del t} + \dot{\w} \sum_{m = 2}^{\infty} m \phi(m) \frac{\del \pw}{\del \phi(m)} \) .
\ee
Demanding $\dot{\Ht}_i \sim 0$ implies the consistency condition
\be
\w \frac{\del \pw}{\del t} + \dot{\w} \sum_{m = 2}^{\infty} m \phi(m) \frac{\del \pw}{\del \phi(m)} \sim f(t) \, ,
\ee
where $f(t)$ is an arbitrary function of time.  We observe that the equation of motion~(\ref{eom}) is invariant under $\pw \gt \pw + g(t)$, where $g(t)$ is an arbitrary function of time, so we are free to apply this transformation to simplify our consistency condition.  If we choose $g(t)$ so that $\w g'(t) = f(t)$, the consistency condition becomes
\be
\w \frac{\del \pw}{\del t} + \dot{\w} \sum_{m = 2}^{\infty} m \phi(m) \frac{\del \pw}{\del \phi(m)} \sim 0 \, .
\ee
By assumption, $\w(t)$ is an invertible function of time, so $\del/\del t = \dot{\w} \, \del/\del \w$.  Our consistency condition can thus be written as
\be
\D \pw \sim 0 \, ,
\ee
where we have defined the operator
\be
\D \equiv \w \frac{\del}{\del \w} + \sum_{m = 2}^{\infty} m \phi(m) \frac{\del}{\del \phi(m)} \, .
\ee
To rule out the possibility of a $\pw$ which satisfies $\D \pw \sim 0$ while $\D \pw \neq 0$, we note that the constraints $\Ht_i$ contain one power of spatial derivatives, while by assumption the scalar momentum $\pw$ is ultralocal.  To satisfy $\D \pw \sim 0$, the quantity $\D \pw$ would need to depend on the constraints $\Ht_i$, and would thus need to contain at least one power of spatial derivatives.  However, applying $\D$ to $\pw$ does not increase the number of spatial derivatives.  It follows that $\D \pw$ cannot contain any spatial derivatives, and thus cannot depend on $\Ht_i$.  The consistency condition can therefore be promoted to
\be
\D \pw = 0 \, .
\ee
To obtain the most general solution to this equation, we first note that $\D \( \w^{-n} \phi(n) \) = 0$, which motivates us to define
\be
\pb(n) \equiv \frac{\phi(n)}{\w^n(t)} \, .
\ee
The most general solution to the condition $\D \pw = 0$ is an arbitrary function of the $\pb(n)$.  The explicit time dependence of $\pw$ is thus determined by its dependence on the phase space variables.

To understand this result, we return briefly to the phase space $(h_{i j}, \pi^{i j})$.  To construct three-scalars out of the tensor $h_{i j}$ and the traceless tensor $\pt^{i j}_T$, we begin by recursively defining $\Pi^{i j}_T(n)$, a chain of $n$ factors of $\pt^{i j}_T$ linked together by factors of $h_{i j}$.  In analogy with our construction of the $\phi(n)$ of equation~(\ref{phin}), we define
\be
\Pi^{i j}_T(0) \equiv h^{i j} = \W^{-1} \Pi^{i j}(0) \, ,
\ee
and
\be
\Pi^{i j}_T(n+1) \equiv \pt^{i a}_T h_{a b} \Pi^{b j}(n) = \w^{-1} \pt^{i a} f_{a b} \Pi^{b j}(n) \, ,
\ee
from which it follows that $\Pi^{i j}_T(n) = \W^{-1} \w^{-n} \Pi^{i j}(n)$.  The contraction $h_{i j}\Pi^{i j}_T(n)$ yields the desired scalars,
\be
\phi_T(n) \equiv h_{i j} \Pi^{i j}_T(n) = \frac{\phi(n)}{\w^n} \, . \label{phint}
\ee
The $\phi_T(n)$ are the only scalars that can be built out of fully connected contractions of $h_{i j}$ and $\pt^{i j}_T$.  In the presence of the constraint $\w \sim \w(t)$, it follows that
\be
\phi_T(n) \sim \pb(n) \, .
\ee
In other words, the $\pb(n)$ are the scalars on the phase space $(\th_{i j}, \pt^{i j})$ which have the correct conformal weight to have been derived from three-scalars on the phase space $(h_{i j}, \pi^{i j})$.  It follows that the $\pb(n)$ are invariant under a rescaling $\w \gt \mu \w$ of the volume factor $\w$, and the condition $\D \pw = 0$ is thus analogous to a renormalization group equation.

\subsection{Summary}
In this section, we developed a formalism for testing when our modified theories of gravity lead to a consistent first class constraint algebra, and hence contain two degrees of freedom.  To develop the formalism, we made the simplifying assumption that the scalar momentum $\pw$ is an ultralocal function of time $t$ and the phase space variables $\th_{i j}$ and $\pt^{i j}$.  This assumption is sufficient to guarantee that the constraints $\Ht_i$ remain first class.  However, for the constraints to be consistent with the equations of motion, $\pw$ must be invariant under renormalization of the volume factor $\w$.  Concretely, $\pw$ must obey the renormalization group equation
\be
\D \pw = 0 \, , \label{rgultralocal}
\ee
where
\be
\D \equiv \w \frac{\del}{\del \w} + \sum_{m = 2}^{\infty} m \phi(m) \frac{\del}{\del \phi(m)} \, .
\ee
Satisfying this equation completely fixes the dependence of $\pw$ on $\w(t)$.  In the next section, we will apply the methods of this section to generalize this result to a more realistic class of scalar momenta.

\section{Realistic Modified Gravity}
\label{realistic}
The ultralocal ansatz has the virtue of simplifying calculations, but it has the defect of being manifestly unphysical: the laws of nature are local, not ultralocal.  In this section, we will apply the formalism developed in the last section to theories in which $\pw$ depends on spatial derivatives of the metric $\th_{i j}$ through a dependence on the Ricci scalar $\Rt$.  Since the $\pgr$ of spatially covariant general relativity belongs to this class, we call it the ``realistic'' class.  As we will demonstrate, realistic $\pw$ must obey stringent consistency conditions in order for the $\Ht_i$ to generate a consistent first class constraint algebra.

\subsection{Constraint Algebra}
\label{rconstraint}
In this section, we will compute $\{ \H_i(x), \H_a(y) \}$ assuming that $\pw$ is a function of $t$, the phase space variables $\th_{i j}$ and $\pt^{i j}$, and the Ricci scalar $\Rt$.  We will then use the result to determine when the constraints $\Ht_i$ remain first class.

As before, we decompose $\Ht_i$ into a tensor part $\J_i \equiv - 2 \th_{i j} \gradt_k \pt^{j k}$ and a scalar part $\K_i \equiv - \w \gradt_i \pw$.  Computing $\{ \H_i(x), \H_a(y) \}$ is then a matter of computing the four brackets in equation~(\ref{hjk}).  The result for $\{ \J_i(x), \J_a(y) \}$ carries over unchanged from equation~(\ref{jj}), but we will have to revisit the brackets involving $\K_i$.  To do so, we will first evaluate the smoothing functional brackets $\{ F_J, G_K \} + \{ F_K, G_J \}$ and $\{ F_K, G_K \}$.  By comparing the ensuing explicit expressions to the formal expressions in equation~(\ref{intjk}), we will derive explicit expressions for the Poisson brackets involving $\K_i$.

Our analysis of the variational derivatives of the smoothing functional $F_K$ defined in equation~(\ref{fjfk}) proceeds exactly as in the ultralocal case up to equation~(\ref{varfk}), where the quantity $\d \pw$ arises.  In this section, we assume that $\pw$ is a function of $t$, $\th_{i j}$, $\pt^{i j}$, and $\Rt$.  To simplify calculations, note that this is equivalent to making $\pw$ a function of $t$, $\Rt$, and the $\phi(n)$ defined in equation~(\ref{phin}).  It follows from this assumption that
\be
\d \pw = \sum_{n = 2}^{\infty} \frac{\del \pw}{\del \phi(n)} \d \phi(n) + \frac{\del \pw}{\del \Rt} \d \Rt \, .
\ee
Substituting this result into equation~(\ref{varfk}), using the identity $\d \Rt = - \Rt^{j k} \d \th_{j k} + \gradt^k \gradt^j \d \th_{j k}$, and integrating by parts yields
\be
\d F_{K} = \w &\int \dcx \( \del_i f^i \) \( \sum_{n = 2}^{\infty} \frac{\del \pw}{\del \phi(n)} \d \phi(n) - \frac{\del \pw}{\del \Rt} \Rt^{j k} \d \th_{j k} \) \nn \\
+ \w &\int \dcx \gradt^j \gradt^k \( \( \del_i f^i \) \frac{\del \pw}{\del \Rt} \) \d \th_{j k} \, .
\ee
Using equation~(\ref{deltaphidelta}), it is now straightforward to compute the variational derivatives of $F_K$,
\be
\frac{\d F_K}{\d \th_{m n}}
= \w &\( \del_i f^i \) \sum_{n = 2}^{\infty} n \frac{\del \pw}{\del \phi(n)} \( \td^{m n}_{j k} \Pi(n)^{j k} - \frac{1}{3} \pt^{m n} \phi(n-1) \) \nn \\
- \w &\( \del_i f^i \) \frac{\del \pw}{\del \Rt} \td_{j k}^{m n} \Rt^{j k} + \w \td_{j k}^{m n} \gradt^j \gradt^k \(\( \del_i f^i \) \frac{\del \pw}{\del \Rt}\) , \nn \\
\frac{\d F_K}{\d \pt^{m n}}
= \w &\( \del_i f^i \) \sum_{n = 2}^{\infty} n \frac{\del \pw}{\del \phi(n)} \td^{j k}_{m n} \Pi(n-1)_{j k} \, . \label{rfkhp}
\ee
The corresponding results for $G_K$ are
\be
\frac{\d G_K}{\d \th_{m n}}
= \w &\( \del_a g^a \) \sum_{m = 2}^{\infty} m \frac{\del \pw}{\del \phi(m)} \( \td^{m n}_{b c} \Pi(m)^{b c} - \frac{1}{3} \pt^{m n} \phi(m-1) \) \nn \\
- \w &\( \del_a g^a \) \frac{\del \pw}{\del \Rt} \td_{b c}^{m n} \Rt^{b c} + \w \td_{b c}^{m n} \gradt^b \gradt^c \(\( \del_a g^a \) \frac{\del \pw}{\del \Rt}\) , \nn \\
\frac{\d G_K}{\d \pt^{m n}}
= \w &\( \del_a g^a \) \sum_{m = 2}^{\infty} m \frac{\del \pw}{\del \phi(m)} \td^{b c}_{m n} \Pi(m-1)_{b c} \, . \label{rgkhp}
\ee
We are now in a position to compute the brackets involving $\K_i$.
\begin{itemize}
\item $\{ \J_i(x), \K_a(y) \}+\{ \K_i(x), \J_a(y) \}$

To compute $\{ \J_i(x), \K_a(y) \}+\{ \K_i(x), \J_a(y) \}$, we first compute $\{ F_J, G_K \} + \{ F_K, G_J \}$.  We begin by substituting equations~(\ref{fjhp}) and~(\ref{rgkhp}) into the bracket $\{ F_J, G_K \}$.  After expanding and simplifying a total derivatives of $\phi(n)$, $\{ F_J, G_K \}$ turns into
\be
\{ F_J, G_K \} = - \w &\int \dcz f^i \( \del_a g^a \) \sum_{m = 2}^{\infty} \frac{\del \pw}{\del \phi(m)} \gradt_i \phi(m) + 2 \w \int \dcz \( \del_a g^a \) \frac{\del \pw}{\del \Rt} \Rt_{i}^{\ k} \gradt_k f^i \nn \\
- 2 \w &\int \dcz \(\gradt_k f^i \) \gradt_i \gradt^k \(\( \del_a g^a \) \frac{\del \pw}{\del \Rt}\) \nn \\
+ \frac{2}{3} \w &\int \dcz \( \del_i f^i \) \gradt_c \gradt^c \(\( \del_a g^a \) \frac{\del \pw}{\del \Rt}\) \nn \\
- \w &\int \dcz \(\del_i f^i \) \( \del_a g^a \) \( \frac{2}{3} \Rt \frac{\del \pw}{\del \Rt} + \sum_{m = 2}^{\infty} m \phi(m) \frac{\del \pw}{\del \phi(m)} \) .
\ee
To finesse this expression, integrate by parts, use the identities $\gradt_i \gradt_j V^i  = \gradt_j \gradt_i V^i + \Rt_{i j} V^i$ and $2 \gradt_j \Rt_{i}^{\ j} = \gradt_i \Rt$, simplify a total derivative of $\pw$, use the identity $\K_i = -\w \gradt_i \pw$, and expand to obtain
\be
\{ F_J, G_K \} = &\int \dcz f^i \K_i \del_a g^a + \frac{4}{3} \w \int \dcz \gradt_k \( \del_i f^i \) \gradt^k \(\( \del_a g^a \) \frac{\del \pw}{\del \Rt}\) \nn \\
- \w &\int \dcz \(\del_i f^i \) \( \del_a g^a \) \( \frac{2}{3} \Rt \frac{\del \pw}{\del \Rt} + \sum_{m = 2}^{\infty} m \phi(m) \frac{\del \pw}{\del \phi(m)} \) . \label{rfjgkforpw}
\ee
Similarly,
\be
\{ F_K, G_J \} = - &\int \dcz g^a \K_a \del_i f^i - \frac{4}{3} \w \int \dcz \gradt_k \( \del_a g^a \) \gradt^k \(\( \del_i f^i \) \frac{\del \pw}{\del \Rt}\) \nn \\
+ \w &\int \dcz \(\del_i f^i \) \( \del_a g^a \) \( \frac{2}{3} \Rt \frac{\del \pw}{\del \Rt} + \sum_{m = 2}^{\infty} m \phi(m) \frac{\del \pw}{\del \phi(m)} \) ,
\ee
so the sum of the two brackets reduces to
\be
\{ F_J, G_K \} + \{ F_K, G_J \} = &\int \dcz f^i \K_i \del_a g^a - \int \dcz g^a \K_a \del_i f^i \nn \\
+ \frac{4}{3} \w &\int \dcz \gradt_k \( \del_i f^i \) \gradt^k \(\( \del_a g^a \) \frac{\del \pw}{\del \Rt}\) \nn \\
- \frac{4}{3} \w &\int \dcz \gradt_k \( \del_a g^a \) \gradt^k \(\( \del_i f^i \) \frac{\del \pw}{\del \Rt}\) .
\ee
After integrating by parts, expanding, and using the identity $\del_i \K_a = \del_a \K_i$, this becomes
\be
\{ F_J, G_K \} + \{ F_K, G_J \} = \int \dcz f^i \K_a \del_i g^a - &\int \dcz g^a \K_i \del_a f^i \nn \\
+\int \dcz \( \del_i f^i \) \( \del_k \del_a g^a \) \M^k - &\int \dcz \( \del_a g^a \) \( \del_k \del_i f^i \) \M^k \, ,
\ee
where
\be
\M_k \equiv - \frac{4}{3} \w \gradt_k \frac{\del \pw}{\del \Rt} \, .
\ee
To extract the bracket $\{ \J_i(x), \K_a(y) \}+\{ \K_i(x), \J_a(y) \}$, integrate by parts, relabel dummy indices, and insert the identities in equation~(\ref{deltaeq}) to yield
\be
\{ F_J , G_K \} + \{ F_K , G_J \} = &\int \dcx \dcy f^i(x) g^a(y) \( \K_a(x) \del_{x^i} \d^3(x-y) - \K_i(y) \del_{y^a} \d^3(x-y) \) \nn \\
+&\int \dcx \dcy f^i(x) g^a(y) \del_{x^i} \( - \M^k(x) \del_{x^k} \del_{x^a} \d^3(x-y) \) \nn \\
- &\int \dcx \dcy f^i(x) g^a(y) \del_{y^a} \( - \M^k(y) \del_{y^k} \del_{y^i} \d^3(x-y) \) .
\ee
By comparing this expression to equation~(\ref{intjk}), it is clear that
\be
\{ \J_i(x),\K_a(y) \} + \{ \K_i(x),\J_a(y) \} = &\K_a(x) \del_{x^i} \d^3(x-y) - \K_i(y) \del_{y^a} \d^3(x-y) \nn \\
+ &\del_{x^i} \( - \M^k(x) \del_{x^k} \del_{x^a} \d^3(x-y) \) \nn \\
- &\del_{y^a} \( - \M^k(y) \del_{y^k} \del_{y^i} \d^3(x-y) \) \, . \label{rjk}
\ee

\item $\{ \K_i(x), \K_a(y) \}$

To compute $\{ \K_i(x), \K_a(y) \}$, we first compute the bracket $\{ F_K, G_K \}$.  Substituting equations~(\ref{rfkhp}) and~(\ref{rgkhp}) into the bracket $\{ F_K, G_K \}$ yields
\be
\{ F_K, G_K \} = \w^2 &\int \dcz \( \del_a g^a \) \frac{\del \pw}{\del \pt^{j k}} \gradt^j \gradt^k \(\( \del_i f^i \) \frac{\del \pw}{\del \Rt}\) \nn \\
- \w^2 &\int \dcz \( \del_i f^i \) \frac{\del \pw}{\del \pt^{j k}} \gradt^j \gradt^k \(\( \del_a g^a \) \frac{\del \pw}{\del \Rt}\) ,
\ee
where
\be
\frac{\del \pw}{\del \pt^{j k}} = \td_{j k}^{b c} \sum_{n = 2}^{\infty} n \frac{\del \pw}{\del \phi(n)} \Pi(n-1)_{b c} \, .
\ee
After integrating by parts and expanding, the bracket becomes
\be
\{ F_K, G_K \} = \int \dcz \( \del_i f^i \) \(\del_k \del_a g^a \) \N^k - \int \dcz \( \del_a g^a \) \( \del_k \del_i f^i \) \N^k, \label{rfkgkforpw}
\ee
where
\be
\N_k \equiv \w^2 \frac{\del \pw}{\del \Rt} \gradt^j \frac{\del \pw}{\del \pt^{j k}} - \w^2 \frac{\del \pw}{\del \pt^{j k}} \gradt^j \frac{\del \pw}{\del \Rt} \, .
\ee
To extract the bracket $\{ \K_i(x), \K_a(y) \}$, integrate by parts, relabel dummy indices, and insert the identities in equation~(\ref{deltaeq}) to obtain
\be
\{ F_K , G_K \} = &\int \dcx \dcy f^i(x) g^a(y) \del_{x^i} \( - \N^k(x) \del_{x^k} \del_{x^a} \d^3(x-y) \) \nn \\
- &\int \dcx \dcy f^i(x) g^a(y) \del_{y^a} \( - \N^k(y) \del_{y^k} \del_{y^i} \d^3(x-y) \) .
\ee
Comparing this expression to equation~(\ref{intjk}), it follows that
\be
\{ \K_i(x), \K_a(y) \} = &\del_{x^i} \( - \N^k(x) \del_{x^k} \del_{x^a} \d^3(x-y) \) \nn \\
- &\del_{y^a} \( - \N^k(y) \del_{y^k} \del_{y^i} \d^3(x-y) \) \, . \label{rkk}
\ee
\end{itemize}
By substituting equations~(\ref{jj}),~(\ref{rjk}), and~(\ref{rkk}) into equation~(\ref{hjk}), and recalling that $\Ht_i = \J_i + \K_i$, we obtain the identity
\be
\{ \Ht_i(x) , \Ht_j(y) \} = &\Ht_j(x) \del_{x^i} \d^3(x-y) - \Ht_i(y) \del_{y^j} \d^3(x-y) \nn \\
+ &\del_{x^i} \( - \I^k(x) \del_{x^k} \del_{x^j} \d^3(x-y) \) - \del_{y^j} \( - \I^k(y) \del_{y^k} \del_{y^i} \d^3(x-y) \) \, ,
\ee
where $\I_k \equiv \M_k + \N_k$, or
\be
\I_k = \w^2 \frac{\del \pw}{\del \Rt} \gradt^j \frac{\del \pw}{\del \pt^{j k}} - \w^2 \frac{\del \pw}{\del \pt^{j k}} \gradt^j \frac{\del \pw}{\del \Rt} - \frac{4}{3} \w \gradt_k \frac{\del \pw}{\del \Rt} \, .
\ee
Expanding the derivatives in this expression and using the fact that $\Ht_i \sim 0$ yields
\be
\{ \Ht_i(x) , \Ht_j(y) \} \sim - &\(\I^k(x)+\I^k(y)\) \del_{x^i} \del_{x^j} \del_{x^k} \d^3(x-y) \nn \\
- & \( \del_{x^i} \I^k(x) \) \del_{x^j} \del_{x^k} \d^3(x-y) \nn \\
+ & \( \del_{y^j} \I^k(y) \) \del_{y^i} \del_{y^k} \d^3(x-y) \, .
\ee
The three terms of this equation are algebraically independent, so the necessary and sufficient condition for the Poisson bracket $\{ \Ht_i(x) , \Ht_j(y) \}$ to vanish is
\be
\I_k \sim 0 \, .
\ee
In the ultralocal case the constraints were automatically first class, but to generate a first class constraint algebra in the realistic case, the scalar momentum $\pw$ must obey the fearsome looking differential equation $\I_k \sim 0$.

As a check, we will now compute the $\I_k$ arising from the $\pgr$ of spatially covariant general relativity.  Recall from equation~(\ref{pigr}) that
\be
\pgr = -\sqrt{\frac{8}{3}} \, \sqrt{ \w^{-2} \phi(2) - \w^{-2/3} \Rt + 2 \L} \, .
\ee
Since $\pgr$ is a function only of $\phi(2)$ and $\Rt$, its partial derivative with respect to $\pt^{i j}$ simplifies,
\be
\frac{\del \pgr}{\del \pt^{j k}} = 2 \frac{\del \pgr}{\del \phi(2)} \pt_{j k} \, .
\ee
After substituting this relation into the definition of $\I_k$ and recalling that $\J_i = - 2 \th_{i j} \gradt_k \pt^{j k}$, $\I_k$ becomes
\be
\I_k(\pgr) = &- \frac{4}{3} \w \gradt_k \frac{\del \pgr}{\del \Rt} - \w^2 \frac{\del \pgr}{\del \Rt} \frac{\del \pgr}{\del \phi(2)} \J_k \nn \\
&+ 2 \w^2 \pt_{j k} \( \frac{\del \pgr}{\del \Rt} \gradt^j \frac{\del \pgr}{\del \phi(2)} - \frac{\del \pgr}{\del \phi(2)} \gradt^j \frac{\del \pgr}{\del \Rt}\) .
\ee
Upon substituting the derivatives
\be
\frac{\del \pgr}{\del \phi(2)} = \frac{4}{3 \w^2} \frac{1}{\pgr} \, , \qquad \frac{\del \pgr}{\del \Rt} = -\frac{4}{3 \w^{2/3}} \frac{1}{\pgr} \, ,
\ee
into $\I_k(\pgr)$, the term in parentheses vanishes.  By using the relations $\K_i = - \w \gradt_i \pw$ and $\H_i = \J_i + \K_i$, we obtain
\be
\I_k(\pgr) = \frac{16}{9 \w^{2/3} \pgr^2} \H_k \, . \label{ikpigr}
\ee
Since $\Ht_i \sim 0$, the scalar momentum $\pgr$ satisfies $\I_k \sim 0$.  The constraints $\Ht_i$ of spatially covariant general relativity thus generate a first class constraint algebra.

\subsection{Consistency of Constraints with Equations of Motion}
In this section, we will compute the time derivative $\dot{\Ht}_i$ for realistic $\pw$ assuming that the constraints $\Ht_i$ are first class, and use the result to determine when the constraints $\Ht_i$ are also preserved by the equations of motion.  The analysis of $\dot{\Ht}_i$ proceeds exactly as in the ultralocal case until we arrive at the expression
\be
\dot{\Ht}_i = - \del_i \(\dot{\w} \pw + \w \frac{\del \pw}{\del t}\) - \dot{\w} \{ \J_i, \Pw \} - \dot{\w} \{ \K_i, \Pw \} \, , \label{rdot}
\ee
where as before
\be
\Pw \equiv \int \dcx \pw \, .
\ee
The point of departure from the ultralocal case is the evaluation of the two Poisson brackets $\{ \J_i, \Pw \}$ and $\{ \K_i, \Pw \}$.  To compute them, we first compute the smoothing functional brackets$\{ F_J, \Pw \}$ and $\{ F_K, \Pw \}$.  As in the ultralocal case, we will obtain brackets involving $\Pw$ by applying the substitution $\del_a g^a \gt \w^{-1}$ to brackets involving $G_K$.
\begin{itemize}
\item $\{ \J_i , \Pw \}$

To obtain the bracket $\{ \J_i, \Pw \}$, we first compute the bracket $\{ F_J, \Pw \}$.  Applying $\del_a g^a \gt \w^{-1}$ to equation~(\ref{rfjgkforpw}) and integrating by parts yields
\be
\{ F_J , \Pw \} = \int \dcx f^i \del_i \(-\pw + \frac{2}{3} \Rt \frac{\del \pw}{\del \Rt} + \sum_{m = 2}^{\infty} m \phi(m) \frac{\del \pw}{\del \phi(m)} -  \w^{-1} \gradt_k \M^k \) ,
\ee
where as before
\be
\M_k = - \frac{4}{3} \w \gradt_k \frac{\del \pw}{\del \Rt} \, .
\ee
It follows from an application of equation~(\ref{fjkpw}) that
\be
\{ \J_i , \Pw \} = \del_i \( - \pw + \frac{2}{3} \Rt \frac{\del \pw}{\del \Rt} + \sum_{m = 2}^{\infty} m \phi(m) \frac{\del \pw}{\del \phi(m)} -  \w^{-1} \gradt_k \M^k \) . \label{rjpw}
\ee

\item $\{ \K_i , \Pw \}$

To compute the bracket $\{ \K_i, \Pw \}$, we first compute the bracket $\{ F_K, \Pw \}$.  After substituting $\del_a g^a \gt \w^{-1}$ and integrating by parts, equation~(\ref{rfkgkforpw}) becomes
\be
\{ F_K, \Pw \} = \int \dcx f^i \del_i \( - \w^{-1} \grad_k \N^k \) ,
\ee
where as before
\be
\N_k = \w^2 \frac{\del \pw}{\del \Rt} \gradt^j \frac{\del \pw}{\del \pt^{j k}} - \w^2 \frac{\del \pw}{\del \pt^{j k}} \gradt^j \frac{\del \pw}{\del \Rt} \, .
\ee
Comparing with equation~(\ref{fjkpw}) yields
\be
\{ \K_i , \Pw \} = \del_i \( - \w^{-1} \grad_k \N^k \) \, . \label{rkpw}
\ee

\end{itemize}
After substituting equations~(\ref{rjpw}) and~(\ref{rkpw}) into the equation of motion~(\ref{rdot}) and recalling that $\I_k = \M_k + \N_k$, we obtain
\be
\dot{\Ht}_i = - \del_i \( \w \frac{\del \pw}{\del t} + \frac{2}{3} \dot{\w} \Rt \frac{\del \pw}{\del \Rt} + \dot{\w} \sum_{m = 2}^{\infty} m \phi(m) \frac{\del \pw}{\del \phi(m)} - \dot{\w} \w^{-1} \gradt_k \I^k \) .
\ee
Since the $\Ht_i$ are assumed to be first class, it follows necessarily that $\I^k \sim 0$.  Demanding $\dot{\Ht}_i \sim 0$ thus implies the consistency condition
\be
\w \frac{\del \pw}{\del t} + \frac{2}{3} \dot{\w} \Rt \frac{\del \pw}{\del \Rt} + \dot{\w} \sum_{m = 2}^{\infty} m \phi(m) \frac{\del \pw}{\del \phi(m)} \sim f(t) \, ,
\ee
where $f(t)$ is an arbitrary function of time.  Recall once again that the equation of motion~(\ref{eom}) is invariant under $\pw \gt \pw + g(t)$, where $g(t)$ is an arbitrary function of time.  By choosing a function $g(t)$ such that $\w g'(t) = f(t)$, the consistency condition becomes
\be
\w \frac{\del \pw}{\del t} + \frac{2}{3} \dot{\w} \Rt \frac{\del \pw}{\del \Rt} + \dot{\w} \sum_{m = 2}^{\infty} m \phi(m) \frac{\del \pw}{\del \phi(m)} \sim 0 \, .
\ee
Since $\w(t)$ is assumed to be an invertible function of time, $\del/\del t = \dot{\w} \, \del/\del \w$.  In analogy with our approach in the ultralocal case, we rewrite the consistency condition as
\be
\D \pw \sim 0 \, ,
\ee
where we have redefined the operator $\D$ as
\be
\D \equiv \w \frac{\del}{\del \w} + \frac{2}{3} \Rt \frac{\del}{\del \Rt} + \sum_{m = 2}^{\infty} m \phi(m) \frac{\del}{\del \phi(m)} \, .
\ee
To rule out the possibility of a $\pw$ which satisfies $\D \pw \sim 0$ while $\D \pw \neq 0$, we note that the constraints $\Ht_i$ contain a term $\gradt_i \pw$, making the constraints higher order in spatial derivatives than $\pw$ itself.  However, by examining a series expansion of $\pw$ in the parameter $\Rt$, one can verify that applying $\D$ to $\pw$ does not alter its order in spatial derivatives.\footnote{Spatial derivatives enter $\pw$ solely through $\Rt$, so the derivative expansion of $\pw$ can be written $\pw = \sum_{k=0}^{\infty} c_k \Rt^k$, where the coefficients $c_k$ depend on $\w$ and the $\phi(n)$.  Applying the $\D$ operator to $\pw$ changes the functional form of the $c_k$, but does not generate higher order powers of $\Rt$.}  It follows that $\D \pw$ cannot depend on $\Ht_i$.  The condition $\D \pw \sim 0$ is therefore equivalent to the apparently stronger condition
\be
\D \pw = 0 \, .
\ee
Since $\D ( \w^{-n} \phi(n) ) = 0$ and $\D ( \w^{-2/3} \Rt ) = 0$, we are led to define the quantities
\be
\pb(n) \equiv \frac{\phi(n)}{\w^n(t)} \, , \qquad \Rb \equiv \frac{\Rt}{\w^{2/3}} \, .
\ee
The most general solution to the condition $\D \pw = 0$ is an arbitrary function of $\Rb$ and the $\pb(n)$.  In this manner, the dependence of $\pw$ on $\w(t)$ is determined by its dependence on the phase space variables.

As before, to understand this result, we return briefly to the phase space $(h_{i j}, \pi^{i j})$.  As shown in section~\ref{ulconsistent}, the only scalars that can be built out of the tensor $h_{i j}$ and the traceless tensor $\pt^{i j}_T$ are the $\phi_T(n) = \w^{-n} \phi(n)$.  If we impose the gauge-fixing constraint $\w \sim \w(t)$, then $\phi_T(n) \sim \pb(n)$; likewise, the Ricci scalar $R$ of the metric $h_{i j}$ obeys $R \sim \Rb$.\footnote{See equation~(\ref{riccigauge}) in appendix C.}  This means that $\Rb$ and the $\pb(n)$ have the correct conformal weight to have been derived from three-scalars on the phase space $(h_{i j}, \pi^{i j})$.  The scalars $\Rb$ and the $\pb(n)$ are thus invariant under a rescaling $\w \gt \mu \w$ of the volume factor $\w$, so once again $\D \pw = 0$ is revealed to be analogous to a renormalization group equation.

As a check, we will now apply the renormalization group equation to the scalar momentum $\pgr$ of spatially covariant general relativity.  Since
\be
\pgr = -\sqrt{\frac{8}{3}} \, \sqrt{ \pb(2) - \Rb + 2 \L} \, ,
\ee
the scalar momentum $\pgr$ satisfies the condition $\D \pgr = 0$; this implies that the constraints of the theory are preserved by the equations of motion.  Combined with the result that $\I_k(\pgr) \sim 0$, which implies that the constraints are also first class, it is now clear within the context of our formalism that the constraints $\Ht_i$ of spatially covariant general relativity generate a consistent first class algebra.  This result justifies the assertions we made in the first paragraph of section~\ref{gr3dof}.

\subsection{Summary}
In this section, we applied the formalism developed in section~\ref{ultralocal} to determine when scalar momenta $\pw$ built out of $\th_{i j}$, $\pt^{i j}$, and $\Rt$ yield a consistent first class constraint algebra.  To ensure the first class character of the constraints $\Ht_i$, it is necessary and sufficient for $\pw$ to obey the condition
\be
\I_k \sim 0 \, ,
\ee
where
\be
\I_k = \w^2 \frac{\del \pw}{\del \Rt} \gradt^j \frac{\del \pw}{\del \pt^{j k}} - \w^2 \frac{\del \pw}{\del \pt^{j k}} \gradt^j \frac{\del \pw}{\del \Rt} - \frac{4}{3} \w \gradt_k \frac{\del \pw}{\del \Rt} \, .
\ee
If $\del \pw/\del \Rt = 0$, then $\I_k = 0$, so ultralocal scalar momenta satisfy this condition trivially.  The scalar momentum $\pgr$ of spatially covariant general relativity depends essentially on $\Rt$, and thus satisfies this condition non-trivially.

To guarantee the preservation of the constraints $\Ht_i$ by the equations of motion, the scalar momentum $\pw$ must also be invariant under renormalization of the volume factor $\w$.  This requires $\pw$ to obey the renormalization group equation
\be
\D \pw = 0 \, ,
\ee
where
\be
\D \equiv \w \frac{\del}{\del \w} + \frac{2}{3} \Rt \frac{\del}{\del \Rt} + \sum_{m = 2}^{\infty} m \phi(m) \frac{\del}{\del \phi(m)} \, .
\ee
This is a generalization of the renormalization group equation~(\ref{rgultralocal}) to include a possible dependence of $\pw$ on $\Rt$.  The scalar momentum $\pgr$ satisfies this condition in addition to the first, so the constraints of spatially covariant general relativity generate a consistent first class constraint algebra.

\section{Conclusions}
In this paper, we developed a general formalism for verifying the consistency of spatially covariant modified theories of the transverse, traceless graviton degrees of freedom.  It was a long road, so it is worth retracing our steps to see the logic of our path.

In section~\ref{gr4}, we showed how to express general relativity as a theory of a spatial metric $h_{i j}$ and its conjugate momentum $\pi^{i j}$.  In this language, the general covariance of the theory is represented on the phase space $(h_{i j}, \pi^{i j})$ by the algebra of the four constraints $\H_{\mu}$.  In section~\ref{gr3}, we showed how to obtain a spatially covariant version of general relativity.  We began in section~\ref{metdec} by splitting the phase space $(h_{i j}, \pi^{i j})$ into the phase space $(\w, \pw)$ of the spatial volume factor and the phase space $(\th_{i j}, \pt^{i j})$ of the unit-determinant metric.  In the context of cosmology on an FRW background, it is natural to represent time diffeomorphism symmetry on the phase space $(\w, \pw)$ and to represent spatial diffeomorphisms on the phase space $(\th_{i j}, \pt^{i j})$; in section~\ref{cosmogauge}, we showed how to achieve this splitting using a cosmological gauge condition.  On an expanding background, $\w$ drops out of the dynamical phase space of the theory, and its conjugate momentum $\pw$ becomes the scalar Hamiltonian density on the phase space $(\th_{i j}, \pt^{i j})$; in section~\ref{solveH0}, we showed how to reduce the phase space by solving the Hamiltonian constraint in cosmological gauge.  By successfully projecting the degrees of freedom of general relativity onto the reduced phase space $(\th_{i j}, \pt^{i j})$, we have shown how to represent the graviton dynamics of general relativity on the class of conformally equivalent spatial metrics.

To modify general relativity, we simply modified the functional form of the scalar momentum $\pw$ while retaining the explicit spatial diffeomorphism symmetry generated by the three constraints $\Ht_i$.  In section~\ref{ultralocal}, we considered the case in which $\pw$ is an ultralocal function of the phase space quantities $\th_{i j}$ and $\pt^{i j}$.  In this case, the consistency of the constraints $\Ht_i$ imposes a single non-trivial condition on the form of $\pw$, namely that it must satisfy a renormalization group equation with flow parameter $\w$.  The renormalization group equation encodes the fact that $\pw$ must be invariant under flow through the space of conformally equivalent spatial metrics.  In section~\ref{realistic}, we applied our formalism to the more realistic case in which $\pw$ is also allowed to depend on $\Rt$, the Ricci scalar of the metric $\th_{i j}$.  In this case, $\pw$ must satisfy a corresponding renormalization group equation, but its form is further constrained by a differential equation that relates its dependence on $\Rt$ to its dependence on the phase space variables $\th_{i j}$ and $\pt^{i j}$.

As a proof of principle, this paper demonstrates the possibility of consistently modifying the graviton equations of motion, but more remains to be done.  In forthcoming work~\cite{godfreytoappear}, we will apply our formalism to search for realistic alternatives to general relativity.  In particular, we will examine scalar momenta $\pw$ with a more general dependence on derivative quantities, such as the Ricci tensor $\Rt_{i j}$ of the unit-determinant metric $\th_{i j}$, which fully determines the spatial curvature in cosmological gauge.  After deriving consistency conditions in the more general case, we will attempt to solve them perturbatively to obtain valid scalar momenta $\pw$ related to the $\pgr$ of general relativity by the deformation of a continuous parameter.  If we discover non-trivial modifications of general relativity that contain only two degrees of freedom, it could open up new lines of theoretical and experimental research.  A null result, on the other hand, would serve as further evidence of the uniqueness of general relativity.  It will be interesting to see just how far we can push this program.

{\it Acknowledgments:}  We thank K.~Hinterbichler and M.~Trodden for helpful discussions.  We also thank the Perimeter Institute for Theoretical Physics in Waterloo, Ontario, where portions of this research were completed.
This work is supported in part by the US Department of Energy DE-AC02-76-ER-03071 (J.K. and G.E.J.M.) and the Alfred P. Sloan Foundation (J.K.).

\newpage
\section*{Appendix A: Covariant Constraint Algebra of GR}
Recall the Poisson bracket of GR
\be
\{ A, B \} \equiv \int \dcz \( \frac{\d A}{\d h_{m n}(z)} \frac{\d B}{\d \pi^{m n}(z)} - \frac{\d A}{\d \pi^{m n}(z)} \frac{\d B}{\d h_{m n}(z)}\)
\ee
and the constraints
\be
\H_0 &\equiv - \sqrt{h} (R - 2\L) + \frac{1}{\sqrt{h}} \( \pi^{ij}\pi_{ij} - \frac{1}{2} (\pi^i_{\ i} )^2 \) \nn \\
\H_i &\equiv -2 h_{ij} \grad_{k} \pi^{jk} \, .
\ee
Our object in this section is to derive the constraint algebra
\be
\{\H_0 (x), \H_0 (y)\}
&= \H^i (x) \del_{x^i} \d^3(x-y) - \H^i (y) \del_{y^i} \d^3(x-y) \nn \\
\{\H_0 (x), \H_i (y)\}
&= \H_0 (y) \del_{x^i} \d^3(x-y) \nn \\
\{\H_i (x), \H_j (y)\}
&= \H_j (x) \del_{x^i} \d^3(x-y) - \H_i (y) \del_{y^j} \d^3(x-y) \, .
\ee
To evaluate these Poisson brackets, we first define the smoothing functionals
\be
F_H \equiv \int \dcx f^0(x) \H_0(x) \, , &\qquad F \equiv \int \dcx f^i(x) \H_i(x) \, , \nn \\
G_H \equiv \int \dcy g^0(y) \H_0(y) \, , &\qquad G \equiv \int \dcy g^a(y) \H_a(y) \, ,
\ee
where the functions $f^0$, $f^i$, $g^0$, and $g^i$ are time-independent smoothing functions.  We then compute the brackets
\be
\{ F_H , G_H \} &= \int \dcx \dcy f^0(x) g^0(y) \{ \H_0(x), \H_0(y) \} \nn \\
\{ F_H , G \} &= \int \dcx \dcy f^0(x) g^a(y) \{ \H_0(x), \H_a(y) \} \nn \\
\{ F , G \} &= \int \dcx \dcy f^i(x) g^a(y) \{ \H_i(x), \H_a(y) \} \, . \label{fghm}
\ee
As in section~\ref{ulca}, we assume that the smoothing functions decay so rapidly that they eliminate all boundary terms generated by integration by parts, but that they are otherwise arbitrary.  This greatly simplifies the explicit evaluation of the brackets of the smoothing functionals.  By comparing the explicit forms of the brackets to the implicit forms in equation~(\ref{fghm}), we will derive explicit formulae for the brackets of the $\H_{\mu}$'s.

To simplify the calculation of the variational derivatives of $F_H$, we will split the Hamiltonian constraint $\H_0$ into a kinetic piece $\H_T$ and a potential piece $\H_V$.  Explicitly, we have $\H_0 = \H_T + \H_V$, where
\be
\H_T &\equiv \frac{1}{\sqrt{h}} \( h_{i k} h_{j l} - \frac{1}{2}h_{ij} h_{kl} \) \pi^{ij}\pi^{kl} \, , \nn \\
\H_V &\equiv - \sqrt{h} (R - 2 \L) \, .
\ee
Similarly, $F_H = F_T + F_V$, where
\be
F_T \equiv \int \dcx f^0(x) \H_T(x) \, , \qquad F_V \equiv \int \dcx f^0(x) \H_V(x) \, .
\ee
Computing the variation $\d F_T$ is straightforward:
\be
\d F_T &= \int \dcx f^0 \( \frac{1}{\sqrt{h}} \( 2 \pi^{i}_{\ k} \pi^{k j} - \pi^k_{\ k} \pi^{i j} \) - \frac{1}{2} \H_T h^{i j} \) \d h_{i j} \nn \\
&+ \int \dcx f^0 \frac{1}{\sqrt{h}} \( 2 \pi_{i j} - h_{i j} \pi^k_{\ k} \) \d \pi^{i j} \, .
\ee
It follows that
\be
\frac{\d F_T}{\d h_{m n}} &= f^0 \( \frac{1}{\sqrt{h}} \( 2 \pi^{m}_{\ k} \pi^{k n} - \pi^k_{\ k} \pi^{m n} \) - \frac{1}{2} \H_T h^{m n} \) \nn \\
\frac{\d F_T}{\d \pi^{m n}} &= f^0 \frac{1}{\sqrt{h}} \( 2 \pi_{m n} - h_{m n} \pi^k_{\ k} \) \, . \label{fthp}
\ee
Likewise,
\be
\frac{\d G_T}{\d h_{m n}} &= g^0 \( \frac{1}{\sqrt{h}} \( 2 \pi^{m}_{\ k} \pi^{k n} - \pi^k_{\ k} \pi^{m n} \) - \frac{1}{2} \H_T h^{m n} \) \nn \\
\frac{\d G_T}{\d \pi^{m n}} &= g^0 \frac{1}{\sqrt{h}} \( 2 \pi_{m n} - h_{m n} \pi^k_{\ k} \) \, . \label{gthp}
\ee

Keeping in mind that $\d R = - \d h_{i j} R^{i j} + \grad^{j}\grad^{i} \d h_{i j} - \grad^{k} \grad_{k} h^{i j} \d h_{i j}$, computing $\d F_V$ is just as straightforward:
\be
\d F_V &= \int \dcx f^0 \( \frac{1}{2} \H_V h^{i j} + \sqrt{h} R^{i j} \) \d h_{i j} \nn \\
&+ \int \dcx \sqrt{h} f^0 \( \grad^{k} \grad_{k} h^{i j} \d h_{i j} - \grad^{j}\grad^{i} \d h_{i j} \) \, .
\ee
Before taking variational derivatives, we exploit our freedom to integrate by parts to pull the covariant derivatives off the metric variation $\d h_{i j}$:
\be
\d F_V &= \int \dcx f^0 \( \frac{1}{2} \H_V h^{i j} + \sqrt{h} R^{i j} \) \d h_{i j} \nn \\
&+ \int \dcx \sqrt{h} \( h^{i j} \grad_{k} \grad^{k} f^0 - \grad^{i} \grad^{j} f^0 \) \d h_{i j} \, .
\ee
It follows that
\be
\frac{\d F_V}{\d h_{m n}} &= f^0 \( \frac{1}{2} \H_V h^{m n} + \sqrt{h} R^{m n} \) + \sqrt{h} \( h^{m n} \grad_{k} \grad^{k} f^0 - \grad^{m} \grad^{n} f^0 \) \nn \\
\frac{\d F_V}{\d \pi^{m n}} &= 0 \, . \label{fvhp}
\ee
Similarly,
\be
\frac{\d G_V}{\d h_{m n}} &= g^0 \( \frac{1}{2} \H_V h^{m n} + \sqrt{h} R^{m n} \) + \sqrt{h} \( h^{m n} \grad_{k} \grad^{k} g^0 - \grad^{m} \grad^{n} g^0 \) \nn \\
\frac{\d G_V}{\d \pi^{m n}} &= 0 \, . \label{gvhp}
\ee

Before computing $\d F$, we integrate by parts inside $F$:
\be
F = 2 \int \dcx h_{i j} \pi^{j k} \grad_{k} f^i \, .
\ee
This simplifies the variational calculation:
\be
\d F = 2 \int \dcx (\grad_{k} f^i) \pi^{j k} \d h_{i j} + 2 \int \dcx (\grad_{k} f^i) h_{i j} \d \pi^{j k} + 2 \int \dcx \pi^{j k} h_{i j} \d \grad_{k} f^i \, .
\ee
To evaluate $\d \grad_{k} f^i$, first expand the covariant derivative as $\grad_{k} f^i = \del_{k} f^i + \G^{i}_{k a} f^a$.  It follows that $ \d \grad_{k} f^i = f^a \d \G^{i}_{k a}$.  The identity
\be
\d \G^{l}_{k i} = \frac{1}{2} h^{l m} \( \grad_i \d h_{k m} + \grad_k \d h_{i m} - \grad_m \d h_{i k} \)
\ee
implies that $2 \pi^{j k} h_{i j} \d \grad_{k} f^i = f^i \pi^{j k} \grad_i \d h_{j k}$.  Substituting this result into the expression for $\d F$ and integrating by parts yields
\be
\d F = 2 \int \dcx (\grad_{k} f^i) \pi^{j k} \d h_{i j} - \int \dcx \grad_i (f^i \pi^{j k}) \d h_{j k} + 2 \int \dcx (\grad_{k} f^i) h_{i j} \d \pi^{j k} \, .
\ee
It follows that
\be
\frac{\d F}{\d h_{m n}} &= 2 (\grad_{k} f^i) \pi^{j k} \d_{i j}^{m n} - \grad_i (f^i \pi^{m n}) \nn \\
\frac{\d F}{\d \pi^{m n}} &= 2 (\grad_{k} f^i) h_{i j} \d^{j k}_{m n} \, . \label{fhp}
\ee
Likewise,
\be
\frac{\d G}{\d h_{m n}} &= 2 (\grad_{c} g^a) \pi^{b c} \d_{a b}^{m n} - \grad_a (g^a \pi^{m n}) \nn \\
\frac{\d G}{\d \pi^{m n}} &= 2 (\grad_{c} g^a) h_{a b} \d^{b c}_{m n} \, . \label{ghp}
\ee
We are now in a position to compute the Poisson brackets of interest.
\begin{itemize}
\item $\{ \H_i(x), \H_j(y) \}$

To calculate $\{ \H_i(x), \H_j(y) \}$, we will first calculate $\{ F, G \}$.  Substituting equations~(\ref{fhp}) and~(\ref{ghp}) into the Poisson bracket yields
\be
\{ F, G \}
&= 2 \int \dcz h_{a b} \pi^{b c} \(\grad_{c} f^i\) \(\grad_{i} g^a\) - 2 \int \dcz h_{i j} \pi^{j k} \(\grad_{k} g^a\) \(\grad_{a} f^i\) \nn \\
&- 2 \int \dcz \(\grad_{c} g^a\) \grad_i \(f^i h_{a b} \pi^{b c}\) + 2 \int \dcz \(\grad_{k} f^i\) \grad_a \(g^a h_{i j} \pi^{j k}\) \, .
\ee
After integrating by parts, applying the identity $\( \grad_i \grad_{j} - \grad_{j} \grad_{i} \) u^a = R^{a}_{\ b i j} u^b$, and recalling that $\H_i = -2 h_{i j} \grad_{k} \pi^{j k}$, this bracket becomes
\be
\{ F, G \}
&= \int \dcz \( f^i \H_a \grad_{i} g^a - g^a \H_i \grad_{a} f^i \) \nn \\
&+ 2 \int \dcz f^i g^a \pi^{j k} \( R_{j i k a} + R_{j a i k} \) .
\ee
It follows from the symmetry ($R_{a b c d} = R_{c d a b}$) and antisymmetry ($R_{a b c d} = - R_{b a c d} = - R_{a b d c}$) properties of the Riemann tensor that $R_{j a i k} = - R_{k i j a}$.  The symmetry property ($\pi^{i j} = \pi^{j i}$) of the momentum tensor then implies that $\pi^{j k} \( R_{j i k a} + R_{j a i k} \) = 0$, so
\be
\{ F, G \}
&= \int \dcz \( f^i \H_a \grad_{i} g^a - g^a \H_i \grad_{a} f^i \) \, .
\ee
Upon expanding the covariant derivatives, the connection terms cancel, yielding
\be
\{ F, G \} = \int \dcz \( f^i \H_a \del_{i} g^a - g^a \H_i \del_{a} f^i \) \, . \label{explicitMM}
\ee
To extract the Poisson brackets $\{ \H_i (x), \H_j(y) \}$, first relabel integration variables,
\be
\{ F, G \} = \int \dcx f^i(x) \H_a(x) \del_{x^i} g^a(x) - \int \dcy g^a(y) \H_i(y) \del_{y^a} f^i(y) \, ,
\ee
then use the identities
\be
g^a(x) = \int \dcy \d^3(x-y) g^a(y) \, , \qquad f^i(y) = \int \dcx \d^3(x-y) f^i(x) \, ,
\ee
to write
\be
\{ F, G \} = \int \dcx \dcy f^i(x) g^a(y) \( \H_a(x) \del_{x^i} \d^3(x-y) - \H_i(y) \del_{y^a} \d^3(x-y) \) .
\ee
By comparing this expression to~(\ref{fghm}), we obtain the identity
\be
\{\H_i (x), \H_j (y)\} = \H_j (x) \del_{x^i} \d^3(x-y) - \H_i (y) \del_{y^j} \d^3(x-y) \, .
\ee

\item$\{\H_0 (x), \H_i (y)\}$

To calculate $\{ \H_0(x), \H_i(y) \}$, we will first calculate $\{ F_H, G \} = \{ F_T, G \} + \{ F_V, G \}$.  Substituting equations~(\ref{fthp}) and~(\ref{ghp}) into the bracket $\{ F_T, G \}$ yields
\be
\{ F_T , G \} = \int \dcz f^0 \grad_a \( g^a \H_T \) .
\ee
Assembling equations~(\ref{fvhp}) and~(\ref{ghp}) into the bracket $\{ F_V, G \}$ yields
\be
\{ F_V , G \} &= \int \dcz f^0 \(\grad_{a} g^a\) \H_V + 2 \int \dcz \sqrt{h} f^0 (\grad_{c} g^a) R_{a}^{\ c} \nn \\
&+ 2 \int \dcz \sqrt{h} \( \grad_{a} g^a \) \grad_{c} \grad^{c} f^0 - 2 \int \dcz \sqrt{h} \(\grad_{c} g^a \) \grad_a \grad^c f^0 \, .
\ee
After integrating the last two terms by parts, the identity $(\grad_a \grad_{c} - \grad_{c} \grad_{a}) g^a = R_{a c} g^a$ implies that
\be
\{ F_V , G \} &= \int \dcz f^0 \( \grad_{a} g^a \) \H_V + 2 \int \dcz \sqrt{h} R_{a}^{\ c} \grad_{c} \( f^0 g^a \) .
\ee
By integrating the last term by parts, using the identity $2 \grad_{c} R_{a}^{\ c} = \grad_a R = \grad_a (R - 2 \L)$, and recalling that $\H_V = \sqrt{h} (2 \L - R)$, the bracket becomes
\be
\{ F_V , G \} = \int \dcz f^0 \grad_{a} \( g^a \H_V \) .
\ee
Combining $\{ F_V , G \}$ with $\{ F_T , G \}$ and recalling that $\H_0 = \H_T + \H_V$ yields
\be
\{ F_H , G \} = \int \dcz f^0 \grad_{a} \( g^a \H_0 \) .
\ee
Since $g^a$ is a three-vector and $\H_0/\sqrt{h}$ is a three-scalar,
\be
\grad_{a} (g^a \H_0) = \del_a \( g^a \H_0 \) ,
\ee
from which it follows that
\be
\{ F_H , G \} = \int \dcz f^0 \del_{a} \( g^a \H_0 \) .
\ee
To extract the bracket $\{ \H_0(x), \H_i(y) \}$, first relabel the variable of integration,
\be
\{ F_H , G \} = \int \dcx f^0(x) \del_{x^a} \( g^a(x) \H_0(x) \) ,
\ee
then use the identity
\be
g^a(x) \H_0(x) = \int \dcy \d^3(x-y) g^a(y) \H_0(y)
\ee
to write
\be
\{ F_T , G \} = \int \dcx \dcy f^0(x) g^a(y) \H_0(y)  \del_{x^a} \d^3(x-y) \, .
\ee
By comparing this expression to~(\ref{fghm}), we obtain the identity
\be
\{\H_0 (x), \H_i (y)\} = \H_0 (y) \del_{x^i} \d^3(x-y) \, .
\ee

\item $\{\H_0 (x), \H_0 (y)\}$

To calculate $\{ \H_0(x), \H_0(y) \}$, we will first calculate
\be
\{ F_H, G_H \} = \{ F_T, G_T \} + \{ F_T, G_V \} + \{ F_V, G_T \} + \{ F_V, G_V \} \, .
\ee
It is straightforward to verify that the brackets $\{ F_T, G_T \}$ and $\{ F_V, G_V \}$ vanish identically. To compute $\{ F_T, G_V \}$, substitute equations~(\ref{fthp}) and~(\ref{gvhp}) into the Poisson bracket to obtain
\be
\{ F_T, G_V \} = 2 &\int \dcz f^0 \pi^{m n} \grad_{m} \grad_{n} g^0 \nn \\ - &\int \dcz f^0 g^0 \frac{1}{\sqrt{h}} \( \frac{1}{2} \H_V \pi^k_{\ k} + 2 \sqrt{h} R_{m n} \pi^{m n} \) .
\ee
Likewise,
\be
\{ F_V, G_T \} = - 2 &\int \dcz g^0 \pi^{m n} \grad_{m} \grad_{n} f^0 \nn \\ + &\int \dcz f^0 g^0 \frac{1}{\sqrt{h}} \( \frac{1}{2} \H_V \pi^k_{\ k} + 2 \sqrt{h} R_{m n} \pi^{m n} \) .
\ee
The sum of the four brackets reduces to
\be
\{ F_H, G_H \} = 2 \int \dcz \( f^0 \pi^{m n} \grad_{m} \grad_{n} g^0 - g^0 \pi^{m n} \grad_{m} \grad_{n} f^0 \) .
\ee
After integrating by parts and recalling that $\H_i = -2 h_{i j} \grad_k \pi^{j k}$, the bracket becomes
\be
\{ F_H, G_H \} = \int \dcz \( f^0 \H^i \grad_i g^0 - g^0 \H^i \grad_i f^0 \) .
\ee
Upon expanding the covariant derivatives in terms of partial derivatives and connection terms, the connection terms cancel to yield
\be
\{ F_H, G_H \} = \int \dcz \( f^0 \H^i \del_i g^0 - g^0 \H^i \del_i f^0 \) .
\ee
To extract the bracket $\{ \H_0(x) , \H_0(y) \}$, relabel integration variables and use the identities
\be
g^0(x) = \int \dcy \d^3(x-y) g^0(y) \, , \qquad f^0(y) = \int \dcx \d^3(x-y) f^0(x)
\ee
to write
\be
\{ F_H, G_H \} = \int \dcx \dcy f^0(x) g^0(y) \( \H^i(x) \del_{x^i} \d^3(x-y) - \H^i(y) \del_{y^i} \d^3(x-y) \) .
\ee
By comparing this expression to~(\ref{fghm}), we obtain the identity
\be
\{ \H_0(x) , \H_0(y) \} = \H^i(x) \del_{x^i} \d^3(x-y) - \H^i(y) \del_{y^i} \d^3(x-y) \, .
\ee
\end{itemize}

\section*{Appendix B: Constraint brackets after imposing $\chi$}
We begin with the four constraints $\H_{\mu}$.  After introducing the gauge-fixing constraint
\be
\chi \equiv \sqrt{h} - \w(t) \, ,
\ee
we need to compute the brackets of each of the five constraints (including $\chi$) with $\chi$.  We introduce the smoothing functionals
\be
F_{\chi} \equiv \int \dcx f_{\chi}(x) \chi(x) \, , \qquad G_{\chi} \equiv \int \dcy g_{\chi}(y) \chi(y) \, ,
\ee
where $f_{\chi}$ and $g_{\chi}$ are arbitrary rapidly-decaying smoothing functions.  We then compute the brackets
\be
\{ F_{\chi} , G_{\chi} \} &= \int \dcx \dcy f_{\chi}(x) g_{\chi}(y) \{\chi (x), \chi (y)\} \nn \\
\{ F_H , G_{\chi} \} &= \int \dcx \dcy f^0(x) g_{\chi}(y) \{\H_0 (x), \chi(y)\} \nn \\
\{ F , G_{\chi} \} &= \int \dcx \dcy f^i(x) g_{\chi}(y) \{\H_i (x), \chi (y)\} \, . \label{fgch}
\ee
The variation $\d F_{\chi}$ is
\be
\d F_{\chi} = \int \dcx f_{\chi} \frac{1}{2} \sqrt{h} h^{i j} \d h_{i j} \, ,
\ee
so
\be
\frac{\d F_{\chi}}{\d h_{m n}} = f_{\chi} \frac{1}{2} \sqrt{h} h^{m n} \, , \qquad \frac{\d F_{\chi}}{\d \pi^{m n}} = 0 \, . \label{df}
\ee
Likewise,
\be
\frac{\d G_{\chi}}{\d h_{m n}} = g_{\chi} \frac{1}{2} \sqrt{h} h^{m n} \, , \qquad \frac{\d G_{\chi}}{\d \pi^{m n}} = 0 \, .
\ee
It follows at once that
\be
\{ F_{\chi}, G_{\chi} \} = 0 \, .
\ee
Comparing with~(\ref{fgch}), we obtain the identity
\be
\{\chi (x), \chi (y)\} = 0 \, .
\ee
We now turn to the brackets of $\chi$ with the $\H_{\mu}$.
\begin{itemize}
\item $\{\H_0 (x), \chi(y)\}$

We split $\{ F_H, G_{\chi} \}$ into $\{ F_H, G_{\chi} \} = \{ F_T, G_{\chi} \} + \{ F_V, G_{\chi} \}$.  Assembling equations~(\ref{fthp}) and~(\ref{df}) into the Poisson bracket $\{ F_T, G_{\chi} \}$ yields
\be
\{ F_T, G_{\chi} \} = \int \dcz f^0 g_{\chi} \frac{1}{2} \pi^k_{\ k} \, .
\ee
The bracket $\{ F_V, G_{\chi} \}$ vanishes identically, so
\be
\{ F_H, G_{\chi} \} = \int \dcx f^0 g_{\chi} \frac{1}{2} \pi^k_{\ k} \, .
\ee
To extract the bracket $\{\H_0 (x), \chi(y)\}$, use the identity
\be
g_{\chi}(x) = \int \dcy g_{\chi}(y) \d^3 (x-y) \, ,
\ee
which yields
\be
\{ F_H, G_{\chi} \} = \int \dcx \dcy f^0(x) g_{\chi}(y) \frac{1}{2} \pi^k_{\ k}(x) \d^3 (x-y).
\ee
Comparing to~(\ref{fgch}), we obtain the identity
\be
\{\H_0 (x), \chi(y)\} = \frac{1}{2} \pi^k_{\ k}(x) \d^3 (x-y).
\ee

\item $\{\H_i (x), \chi(y)\}$

From equation~(\ref{fhp}), it follows that
\be
\{ F, G_{\chi} \} = - \int \dcz g_{\chi} \sqrt{h} \grad_{i} f^i \, .
\ee
Integrating by parts and using the fact that $g_{\chi}$ is a scalar, this bracket becomes
\be
\{ F, G_{\chi} \} = \int \dcx f^i \sqrt{h} \del_{i} g_{\chi} \, .
\ee
To extract the bracket $\{\H_i (x), \chi(y)\}$, use the identity
\be
g_{\chi}(x) = \int \dcy g_{\chi}(y) \d^3 (x-y)
\ee
to write
\be
\{ F, G_{\chi} \} = \int \dcx \dcy f^i(x) g_{\chi}(y) \sqrt{h(x)} \del_{x^i} \d^3 (x-y) \, .
\ee
Comparing to~(\ref{fgch}), we obtain the identity
\be
\{\H_i (x), \chi (y)\} = \sqrt{h(x)} \del_{x^i} \d^3 (x-y) \, .
\ee

\end{itemize}

\section*{Appendix C: Conformal Decomposition}
Consider a metric $g_{\mu \nu}$ in a number of dimensions $d$.  Denote the determinant of $g_{\mu \nu}$ by $g$.  Define the positive conformal factor
\be
\W \equiv |g|^{1/d} > 0
\ee
and the metric
\be
\tg_{\mu \nu} \equiv |g|^{-1/d} g_{\mu \nu}
\ee
so that
\be
g_{\mu \nu} = \W \tg_{\mu \nu} \, .
\ee
By construction, the signature of $\tg_{\mu \nu}$ is the same as that of $g_{\mu \nu}$.  Denote the determinant of $\tg_{\mu \nu}$ by $\tg$.  From the definition of $\tg_{\mu \nu}$, it follows that $\tg = g/|g|$, so $\tg = \pm 1$, depending on the signature of $g_{\mu \nu}$.  We therefore call $\tg_{\mu \nu}$ a unit-determinant metric.

The inverse metrics are related by $g^{\mu \nu} = \tg^{\mu \nu} \W^{-1}$.  We denote the covariant derivative with respect to $g_{\mu \nu}$ by $\grad_{\mu}$, and the covariant derivative with respect to $\tg_{\mu \nu}$ by $\gradt_{\mu}$.  The connection $\G^{\l}_{\mu \nu}$ defined by $g_{\mu \nu}$ is
\be
\G^{\l}_{\mu \nu} = \frac{1}{2} g^{\l \s} \( \del_{\mu} g_{\nu \s} + \del_{\nu} g_{\mu \s} - \del_{\s} g_{\mu \nu} \) ,
\ee
while the connection $\Gt^{\l}_{\mu \nu}$ defined by $\tg_{\mu \nu}$ is
\be
\Gt^{\l}_{\mu \nu} = \frac{1}{2} \tg^{\l \s} \( \del_{\mu} \tg_{\nu \s} + \del_{\nu} \tg_{\mu \s} - \del_{\s} \tg_{\mu \nu} \) .
\ee
The connection $\Gt^{\l}_{\mu \nu}$ obeys $\Gt^{\l}_{\mu \nu} = \G^{\l}_{\mu \nu} - C^{\l}_{\mu \nu}$, where
\be
C^{\l}_{\mu \nu} = \( \d_{\mu \nu}^{\l \s} - \frac{1}{2} \tg^{\l \s} \tg_{\mu \nu} \) \del_{\s} \log \W \, .
\ee
For convenience, we can write $\W$ in terms of a scalar field $\f$ and a constant $\W_0$ as
\be
\W \equiv \W_0 e^{2 \f} \, ,
\ee
in which case
\be
C^{\l}_{\mu \nu} = \d_{\mu}^{\l} \gradt_{\nu} \f + \d_{\nu}^{\l} \gradt_{\mu} \f - \tg_{\mu \nu} \gradt^{\l} \f \, .
\ee
The Riemann tensor of $g_{\mu \nu}$ is
\be
R^{\l}_{\ \k \mu \nu} = \del_{\mu} \G^{\l}_{\k \nu} - \del_{\nu} \G^{\l}_{\k \mu} + \G^{\l}_{\mu \s} \G^{\s}_{\k \nu} - \G^{\l}_{\nu \s} \G^{\s}_{\k \mu} \, ,
\ee
while the Riemann tensor of $\tg_{\mu \nu}$ is
\be
\Rt^{\l}_{\ \k \mu \nu} = \del_{\mu} \Gt^{\l}_{\k \nu} - \del_{\nu} \Gt^{\l}_{\k \mu} + \Gt^{\l}_{\mu \s} \Gt^{\s}_{\k \nu} - \Gt^{\l}_{\nu \s} \Gt^{\s}_{\k \mu} \, .
\ee
Using $\G^{\l}_{\mu \nu} = \Gt^{\l}_{\mu \nu} + C^{\l}_{\mu \nu}$, the Riemann tensor $R^{\l}_{\ \k \mu \nu}$ can be rewritten as
\be
R^{\l}_{\ \k \mu \nu} = \Rt^{\l}_{\ \k \mu \nu} + C^{\l}_{\mu \s} C^{\s}_{\k \nu} - C^{\l}_{\nu \s} C^{\s}_{\k \mu} &+\del_{\mu} C^{\l}_{\k \nu} + \Gt^{\l}_{\mu \s} C^{\s}_{\k \nu} - \Gt^{\s}_{\mu \k} C^{\l}_{\s \nu} \nn \\
&- \del_{\nu} C^{\l}_{\k \mu} - \Gt^{\l}_{\nu \s} C^{\s}_{\k \mu} + \Gt^{\s}_{\nu \k} C^{\l}_{\mu \s} \, .
\ee
Using
\be
\gradt_{\mu}C^{\l}_{\k \nu} - \gradt_{\nu}C^{\l}_{\k \mu} &= \del_{\mu} C^{\l}_{\k \nu} + \Gt^{\l}_{\mu \s} C^{\s}_{\k \nu} - \Gt^{\s}_{\mu \k} C^{\l}_{\s \nu} \nn \\
&- \del_{\nu} C^{\l}_{\k \mu} - \Gt^{\l}_{\nu \s} C^{\s}_{\k \mu} + \Gt^{\s}_{\nu \k} C^{\l}_{\s \mu} \, ,
\ee
$R^{\l}_{\ \k \mu \nu}$ becomes
\be
R^{\l}_{\ \k \mu \nu} = \Rt^{\l}_{\ \k \mu \nu} + C^{\l}_{\mu \s} C^{\s}_{\k \nu} - C^{\l}_{\nu \s} C^{\s}_{\k \mu} + \gradt_{\mu}C^{\l}_{\k \nu} - \gradt_{\nu}C^{\l}_{\k \mu} \, . \label{riemann}
\ee
The Ricci tensor of $g_{\mu \nu}$ is $R_{\mu \nu} = R^{\l}_{\ \mu \l \nu}$; the Ricci tensor of $\tg_{\mu \nu}$ is $\Rt_{\mu \nu} = \Rt^{\l}_{\ \mu \l \nu}$.  Tracing equation~(\ref{riemann}) appropriately yields
\be
R_{\mu \nu} = \Rt_{\mu \nu} + C^{\l}_{\l \s} C^{\s}_{\mu \nu} - C^{\l}_{\nu \s} C^{\s}_{\mu \l} + \gradt_{\l}C^{\l}_{\mu \nu} - \gradt_{\nu}C^{\l}_{\mu \l} \, .
\ee
We now express $R_{\mu \nu}$ in terms of $\Rt_{\mu \nu}$ and derivatives of $\f$.  Recalling that $\d^{\mu}_{\mu} = d$, we find
\be
C^{\l}_{\l \s} &= d \gradt_{\s} \f \nn \\
C^{\l}_{\mu \s} C^{\s}_{\nu \l} &= (d+2) (\gradt_{\mu} \f) (\gradt_{\nu} \f) - 2 \tg_{\mu \nu}(\gradt_{\a} \f) (\gradt^{\a} \f) \, ,
\ee
so
\be
R_{\mu \nu} = \Rt_{\mu \nu} &+ (d-2)  (\gradt_{\mu} \f) (\gradt_{\nu} \f) - (d-2) \tg_{\mu \nu} (\gradt_{\s} \f) (\gradt^{\s} \f) \nn \\
&- (d-2) \gradt_{\mu} \gradt_{\nu} \f - \tg_{\mu \nu} \gradt_{\s} \gradt^{\s} \f \, . \label{riccit}
\ee
The Ricci scalar for $g_{\mu \nu}$ is $R = g^{\mu \nu} R_{\mu \nu}$; the Ricci scalar for $\tg_{\mu \nu}$ is $\Rt = \tg^{\mu \nu} \Rt_{\mu \nu}$.  In terms of covariant derivatives of $\f$, we have
\be
\W R = \Rt - (d-1)(d-2) (\gradt_{\a} \f) (\gradt^{\a} \f) - 2 (d-1) \gradt^{\s} \gradt_{\s} \f \, . \label{riccis}
\ee

In three dimensions, the Weyl tensor vanishes, so the Riemann tensor is completely determined by the Ricci tensor and the metric via
\be
R_{l k m n} = &\frac{1}{d-2} \( g_{l m} R_{k n} - g_{l n} R_{k m} - g_{k m} R_{l n} + g_{k n} R_{l m} \) \nn \\
- &\frac{1}{(d-1)(d-2)} (g_{l m} g_{k n}-g_{l n} g_{k m}) R \, . \label{3dr}
\ee
In this case, it suffices to compute the Ricci tensor.  When $d=3$, our previous formulas reduce to
\be
R_{i j} &= \Rt_{i j} + (\gradt_{i} \f) (\gradt_{j} \f) - \tg_{i j} (\gradt_{k} \f) (\gradt^{k} \f) - \gradt_{i} \gradt_{j} \f - \th_{i j} \gradt_{k} \gradt^{k} \f \, , \nn \\
\W R &= \Rt - 2 (\gradt_{k} \f) (\gradt^{k} \f) - 4 \gradt^{k} \gradt_{k} \f \, .
\ee
The condition $\w \sim \w(t)$ amounts to $\f \sim \f(t)$, so in cosmological gauge we have
\be
R_{i j} \sim \Rt_{i j} \, , \qquad \W R \sim \Rt \, . \label{riccigauge}
\ee

\end{document}